\def\smalllinestretch{\renewcommand{\baselinestretch}{0.7}}
\def\largelinestretch{\renewcommand{\baselinestretch}{1.0}}
 \def\Tr {\, \mbox{Tr} \,}
 \def\tr {\, \mbox{tr} \,}
 \def\Trp{\, \mbox{Tr}^{\prime} \,}
 \def\trp{\, \mbox{tr}^{\prime} \,}

 \def\al{{\alpha}}
 \def\be{{\beta}}
 \def\U{\Phi}

\def\square{\hbox{\vrule\vbox{\hrule\phantom{o}\hrule}\vrule}}

\documentstyle{article}

\def\largelinestretch{\renewcommand{\baselinestretch}{1.2}}
\largelinestretch\small\normalsize
\title{
        Reduction of Vector and Axial--Vector Fields in a Bosonized
        Nambu--Jona--Lasinio Model
      }
 \author{
A.A.Bel'kov${}^1$,
A.V.Lanyov${}^1$,
A.Schaale${}^2$
\\
\\
\small
${}^1$
        Particle Physics Laboratory, Joint Institute for Nuclear Research,
\hfill\\
\small
        Head Post Office, P.O. BOX 79, 101000 Moscow, Russia
\hfill\\
\small
${}^2$
        DESY--IfH Zeuthen, Postfach 15735 Zeuthen, Germany
\hfill\\
}
\begin{document}
\largelinestretch\normalsize
   \thispagestyle{empty}
   \begin{titlepage}
   \thispagestyle{empty}
   \maketitle
   \begin{abstract}

    We derive the effective action for pseudoscalar mesons by
integrating out vector and axial--vector collective fields in the
generating functional of the bosonized NJL--model.
    The corresponding modifications of the nonlinear effective
Lagrangian and the bosonized currents, arising at $O(p^4)$, are
discussed.

   \end{abstract}
   \end{titlepage}
   %

%---------------------------------------------------------------------------
%
\section*{1. Introduction}
%
%---------------------------------------------------------------------------

    A renewal of interest in chiral Lagrangian theory was
excited by recent progress in the construction of realistic
effective chiral meson Lagrangians including higher order derivative
terms as well as the gauge Wess--Zumino term from low--energy
approximations of QCD.
    The program of bosonization of QCD, which was started about 20
years ago, in the strong sense is of course also beyond our present
possibilities.
    Nevertheless there is some success related to the application of
functional methods to approximate forms of QCD (see
\cite{kleinert}-\cite{ball} and references therein) or to
QCD--motivated effective quark models \cite{kikkawa}-\cite{rafael}
which are extensions of the well--known Nambu--Jona--Lasinio (NJL)
model \cite{njl}.
    These functional methods can be applied also to the bosonization of
the effective four--quark nonleptonic weak and electromagnetic--weak
interactions with strangeness change $|\Delta S|=1$ by using the
generating functional for Green functions of quark currents introduced
in \cite{pich-rafael}, \cite{bosoniz-our}.

    The NJL model, which we consider in this paper, incorporates not
only all relevant symmetries of the quark flavour dynamics of
low--energy QCD, but also offers a simple scheme of the spontaneous
breakdown of chiral symmetry arising from the explicit symmetry
breaking terms due to the quark masses.
    In this scheme the current quarks transit into constituent ones
due to the appearance of a nonvanishing quark condensate, and
light composite pseudoscalar Nambu--Goldstone bosons emerge
accompanied also by heavier dynamical vector and axial--vector mesons
with correct relative weights arising from renormalization.
    The composite vector and axial--vector resonances are naturally
required if one wants to describe the low energy aspects of QCD in
a wider energy range up to typical masses of O(1GeV).
    (For other approaches to the problem of introducing chiral
couplings for vector and axial--vector mesons to Goldstone bosons
into the effective chiral Lagrangian, see for example
refs.\cite{resonans}-\cite{donoghue-strongres} and references
therein.)

    Independently from the method of including the vector and
axial--vector fields in the effective chiral Lagrangian, integrating
out the heavy meson resonances essentially modifies the coupling
constants of the pseudoscalar low--energy interactions.
    In particular, in refs.\cite{ecker-strongres},
\cite{donoghue-strongres} it was shown that the structure
constants $L_i$ of the Gasser--Leutwyler general expression for the
$O(p^4)$ pseudoscalar Lagrangian \cite{Gasser-Leutwyler} are largely
saturated by the resonance exchange contributions giving a product of
terms of $O(p^2)$.
    But in this case, if the $O(p^4)$ Lagrangian contains meson
resonances, their elimination can lead to the double counting mentioned in
ref.\cite{ecker-strongres}.
    The resonance contributions to the purely pseudoscalar chiral weak
Lagrangian and the modification of its structure, induced by integrating
out the heavy meson exchanges, were discussed recently in
refs.\cite{ecker-weakres},\cite{isidori-weakres} in both the frame of the
factorization approximation and in the weak deformation approach.

    In this paper we consider the effective nonlinear Lagrangian for
pseudoscalar mesons which arises after integrating out the explicit
vector and axial--vector resonances in the generating functional of
the bosonized NJL model.
    To perform such integration we use a method based on the
invariance of the modulus of the quark determinant under a chiral
transformations and on the application of the static equations of
motion to a special configuration of the chiral rotated fields.
    The elimination of vector and axial--vector degrees of freedom
from the modulus of the quark determinant leads to a modification of the
general structure of the effective strong Lagrangian for the
pseudoscalar sector at $O(p^4)$ and to a redefinition of the
corresponding Gasser-Leutwyler structure coefficients $L_i$.
    This method of  reduction of meson resonances can be extended to
the procedure \cite{bosoniz-our} of chiral  bosonization of weak
and electromagnetic--weak currents and can be used for obtaining the
corresponding reduced meson currents entering to the bosonized
nonleptonic weak Lagrangians.

    In such approximation the problem of double counting does not
arise.
    The effect of $\pi A_1$--mixing, being most important for the
description of radiative weak decays, is taken into account by the
corresponding $\pi A_1$--diagonalization factor.
    This factor appears explicitly in the bosonized strong
Lagrangians and weak and electromagnetic--weak currents after
reduction of the vector and axial--vector fields.
    With such modification of the strong Lagrangian and currents it is
possible to reproduce within the nonlinear parameterization of chiral
symmetry most of the results of the linear Lagrangian approach
\cite{SJPN-volkov} concerning $\pi A_1$--mixing effects and meson
resonance exchange contributions.

    In Section 1 we discuss the basic formalism and display all
definitions and constants related to the bosonization of quarks in
NJL model.
    For convenience, all cumbersome expressions resulting from the
heat--kernel computation of the quark determinant are given in the
Appendix A.
    The total expressions for the bosonized effective Lagrangians
including vector and axial--vector fields are presented in the
Appendix B up to $O(p^4)$ terms.
    In Section 3 we consider the static equations of motion for
chiral rotated collective meson fields in unitary gauge.
    Applying these equations of motion we eliminate the heavy meson
resonances from the modulus of the quark determinant and obtain in
such a way the effective pseudoscalar strong Lagrangian with reduced
vector and axial--vector degrees of freedom.
    The reduced pseudoscalar $(V-A)$ and $(S-P)$ currents corresponding to
the respective quark currents and quark densities are obtained in
Section 4.
    In Section 5 we discuss the results of some numerical estimations
and phenomenological analysis of the structure constants for the
reduced strong Lagrangian and currents.

%---------------------------------------------------------------------------
%
\section*{2. Bosonization of the NJL model}
%
%---------------------------------------------------------------------------

    The starting point of our consideration is the NJL Lagrangian of
the effective four--quark interaction which has the form \cite{njl}:
\begin{equation}
{\cal L}_{NJL} = \overline{q}(i\widehat{\partial }-m_{0})q + {\cal L}_{int}
\label{njl-lagr}
\end{equation}
with
$$
{\cal L}_{int} = 2G_{1}\bigg\{
   \bigg(\overline{q}\frac{\lambda^{a}}{2}q \bigg)^{2}
  +\bigg(\overline{q}i\gamma^{5}\frac{\lambda^{a}}{2}q \bigg)^{2}
                       \bigg\}
                -2G_{2}\bigg\{
   \bigg(\overline{q}\gamma^\mu \frac{\lambda^{a}}{2}q\bigg)^{2}
  +\bigg(\overline{q}\gamma^\mu \gamma^{5}\frac{\lambda^{a}}{2}q\bigg)^{2}
                       \bigg\}\,.
$$
    Here $G_{1}$ and $G_{2}$ are some universal coupling constants;
$m_0=diag(m_0^1,m_0^2, ...,m_0^n)$ is the current quark mass matrix
(summation over repeated indices is assumed), and $\lambda^a$ are the
generators of the $SU(n)$ flavour group normalized according to
$\tr \lambda^{a} \lambda^{b} = 2\delta_{ab}$.
    Using a standard quark bosonization approach based on path
integral techniques one can derive an effective meson
action from the NJL Lagrangian (\ref{njl-lagr}).
   First one has to introduce collective fields for the scalar $(S)$,
pseudoscalar $(P)$, vector $(V)$ and axial--vector $(A)$ colorless mesons
associated to the following quark bilinears:
$$
   S^{a}=-4G_{1}\overline{q}\frac{\lambda^{a}}{2}q\,,\enspace
   P^{a}=-4G_{1}\overline{q}i\gamma^{5}\frac{\lambda^{a}}{2}q\,, \enspace
   V_\mu^{a}=i4G_{2}\overline{q}\gamma_\mu \frac{\lambda^{a}}{2}q\,,\enspace
   A_\mu^{a}=i4G_{2}\overline{q}\gamma_\mu \gamma^{5}\frac{\lambda^{a}}{2}q\,.
$$
    After substituting these expressions into ${\cal L}_{NJL}$ the
interaction part of the Lagrangian is of Yukawa form.
    The part of ${\cal L}_{NJL}$ which is bilinear in the quark fields
can be rewritten as
$$
{\cal L} =  \overline{q} i{\bf \widehat{D}} q
$$
with ${\bf \widehat{D}}$ being the Dirac operator:
\begin{eqnarray}
 i{\bf \widehat{D}} &=& i(\widehat{\partial}+\widehat{V}+\widehat{A}\gamma^5)
                      - P_{R}(\Phi +m_{0}) - P_{L}(\Phi^{+}+m_{0})
\nonumber
\\
        &=& [i(\widehat{\partial} +\widehat{A}_R) - (\Phi    +m_0)] P_R
           +[i(\widehat{\partial} +\widehat{A}_L) - (\Phi^{+}+m_0)] P_L.
\label{dirac}
\end{eqnarray}
    Here $\Phi = S + iP$, $\widehat{V} = V_{\mu} \gamma^{\mu}$,
$\widehat{A} = A_{\mu} \gamma^{\mu}$;
$P_{L,R}={1 \over 2}(1 \pm \gamma_5)$ are chiral projectors;
$\widehat{A}_{L,R} = \widehat{V} \pm \widehat{A}$ are right and left
combinations of fields, and
$$
    S=S^a\frac{\lambda^a}{2} , \quad
    P=P^a\frac{\lambda^a}{2} , \quad
    V_{\mu}=-iV_{\mu}^a \frac{\lambda^a}{2} ,\quad
    A_{\mu}=-iA_{\mu}^a \frac{\lambda^a}{2}
$$
are the matrix-valued collective fields.

    After integration over quark fields the generating
functional, corresponding to the effective action of the NJL model for
collective meson fields, can be presented in the following form:
\begin{equation}
{\cal Z} = \int {\cal D}\Phi\,{\cal D}\Phi^+\,{\cal D}V\,{\cal D}A
           \,\,\mbox{exp}\big[ i{\cal S}(\Phi,\Phi^+,V,A)\big]\,,
\label{genfunc}
\end{equation}
where
\begin{equation}
{\cal S}(\Phi,\Phi^+,V,A) =
          \int d^4 x
          \bigg [ - {1 \over{4G_1}} \mbox{tr} [(\Phi - m_0)^+(\Phi-m_0)]
           - \, {1 \over{4G_2}} \mbox{tr} (V_\mu^2 + A_\mu^2)\bigg ]
           -i\,\Trp [ \log( i \widehat{\bf D} ) ]
\label{action}
\end{equation}
is the effective action for scalar, pseudoscalar, vector and
axial--vector mesons.
    The first term in (\ref{action}), quadratic in meson fields,
arises from the linearization of the four--quark interaction.
    The second term is the quark determinant describing the
interaction of mesons.
    The trace $\Trp$ is to be understood as a space--time integration
and a ``normal'' trace over Dirac, colour and flavour indices:
$$\Trp = \int d^4 x \Tr \,,\quad
  \Tr  = \tr_{\gamma} \cdot \tr_c \cdot \tr_f \,.
$$
    The quark determinant can be evaluated either by expansion in
quark loops \cite{kikkawa}-\cite{volkov-ebert} or by the heat-kernel
technique with proper--time regularization \cite{heat-other}.
    Then, the real part of $\log \big(\det i{\bf \widehat{D}}\big)$
contributes to the non--anomalous part of the effective Lagrangian
while the imaginary part of it gives the anomalous effective
Lagrangian of Wess and Zumino which is related to chiral anomalies
\cite{WZ}.

    The modulus of the quark determinant is presented in the
heat--kernel method as the expansion over the so-called Seeley--deWitt
coefficients $h_k$:
\begin{equation}
\log |\det i{\bf \widehat{D}}| \,  =
 -\, {1\over 2} \frac{\mu^4}{(4 \pi)^2} \sum \limits_{k}
                \frac{\Gamma (k-2, \mu^2 / \Lambda^2)}{\mu^{2k}}
                \Trp h_k,
\label{logarithm}
\end{equation}
where
$$
\Gamma (\alpha,x)=\int^{\infty }_{x} d t \, e^{-t}t^{\alpha-1}
$$
is incomplete gamma function; $\mu$ plays the role of some empirical
mass scale parameter which will fix the regularization in the region
of low momenta, and $\Lambda$ is the intrinsic regularization cutoff
parameter.
    There are some technical reasons for using the heat
kernel method instead of the ``straight'' method of calculating quark
loops.
    The main advantage of this method is that its recursive algorithms
can be adopted on Computer Algebra Systems such as FORM or REDUCE
quite effectively.
    In particular higher order derivative contributions can
conveniently be calculated, too.
    The formulae for the Seeley--deWitt coefficients $h_k$ up to $k=6$
obtained in \cite{heat-our} and the full expressions for $p^4$-- and
$p^6$--contributions to the bosonized meson Lagrangian are presented in
the Appendices A,B.

    We will consider a nonlinear parameterization of chiral symmetry
corresponding to the following representation of $\Phi$:
\begin{eqnarray*}
               \Phi = \Omega \,\Sigma \,\Omega ,
\end{eqnarray*}
where $\Sigma(x)$ is the matrix of scalar fields belonging to the
diagonal flavour group while matrix $\Omega(x)$ represents the
pseudoscalar degrees of freedom $\varphi$ living in the coset space
$U(n)_L \! \times \! U(n)_R/U_V(n)$, which can be parameterized by
the unitary matrix
\begin{eqnarray*}
  \Omega (x) = \exp \left(\frac{i}{\sqrt{2}F_0} \varphi (x) \right)\,,
\quad
  \varphi (x) = \varphi^a(x)\frac{\lambda^a}{2}\,,
\end{eqnarray*}
with $F_0$ being the bare $\pi$ decay constant.
    Under chiral rotations
\begin{eqnarray*}
q \rightarrow \widetilde{q} = \left ( P_L \xi_L + P_R \xi_R \right ) q
\end{eqnarray*}
the fields $\Phi$ and $A^\mu_{R/L}$ are transformed as
\begin{eqnarray*}
     \Phi  \rightarrow \widetilde{\Phi} = \xi_L \Phi \xi^+_R
\end{eqnarray*}
and
\begin{eqnarray}
A^\mu_R \rightarrow \widetilde{A}^\mu_R =
        \xi_R ( \partial^\mu + V^\mu + A^\mu ) \xi^+_R, \quad
A^\mu_L \rightarrow \widetilde{A}^\mu_L =
        \xi_L  ( \partial^\mu + V^\mu - A^\mu ) \xi^+_L.
\label{rotated-fields}
\end{eqnarray}
    For the unitary gauge $\xi^+_L = \xi_R = \Omega$ the rotated Dirac
operator (\ref{dirac}) gets the form
\begin{eqnarray}
i{\bf \widehat{D}} \rightarrow i{\widehat{\widetilde{\bf D}}}
  =(P_L \Omega + P_R \Omega^+)i{\bf \widehat{D}}(P_L \Omega + P_R \Omega^+)
  = i\big(\widehat{\partial} + \widehat{\widetilde{V}}
   +\widehat{\widetilde{A}}\gamma_5 \big) - \Sigma.
\label{eq4}
\end{eqnarray}
    It is worth noting that under local $U_L(n) \times U_R(n)$
transformations the modulus of the quark determinant is invariant,
while the quadratic terms of $V_{\mu}, \, A_{\mu}$ and the chiral
anomaly do not respect this invariance.

    Note that there arises a quark condensate
$<\!\! \overline{q}q \!\!> \neq 0$ owing to the nonvanishing vacuum
expectation value of the scalar meson field $S_0$ realizing
spontaneous breakdown of chiral symmetry.
    In fact, assuming approximate flavor symmetry of the condensate
and using the equation of motion
$S_0 = -2G_1 \sqrt{\frac{2}{n} \overline{q}q}$ one gets
$$
  <\!\!S_0\!\!> = -2G_1 \sqrt{\frac{2}{n}}<\!\! \overline{q}q \!\!>
                \equiv i2G_1 \sqrt{\frac{2}{n}}
                \Tr \{ {i{\bf \widehat{D}}}(\Phi=\mu, V=A=0) \}^{-1}\,.
$$
    Thus, our mass scale $\mu = \frac{1}{\sqrt{2n}}<\!\!S_0\!\!>$ is
determined by $G_1$ and $<\!\! \overline{q}q \!\!>$ or (using the
explicit expression for the condensate by a loop integral) by $G_1$,
the cutoff $\Lambda$ and the current mass $m_0$.

    Taking into account the equations of motion for nonrotated scalar
and pseudoscalar meson fields in nonlinear parameterization one can
derive from (\ref{action}) and eqs.(\ref{append1},\ref{append2}) of
the Appendix B the following general expression of the effective meson
Lagrangian including $p^2$-- and $p^4$--interactions:
\begin{eqnarray}
{\cal L}^{(nred)}_{eff}&=&-\frac{F^2_0}{4} \tr \big( L_{\mu} L^{\mu} \big)
                          +\frac{F_0^2}{4} \tr \big( MU + U^+M \big)
\nonumber
\\
&&+ \bigg( L_1-\frac{1}{2}L_2 \bigg)\,\big( \tr L_{\mu} L^{\mu}\big)^2
  + L_2 \tr \bigg(\frac{1}{2}[L_{\mu},L_{\nu}]^2+3(L_{\mu}L^{\mu})^2\bigg)
  + L_3 \tr \big( (L_{\mu} L^{\mu})^2 \big)
\nonumber
\\
&&+ L_4 \tr \big( D_{\mu}U\,\overline{D}^{\mu}U^+ \big)\,
        \tr M\big( U + U^+ \big)
  + L_5 \tr D_{\mu}U\,\overline{D}^{\mu}U^+\big( MU + U^+M \big)
\nonumber
\\
&&+ L_6\,\Big( \tr \big( MU + U^+M \big) \Big)^2
  + L_7\,\Big( \tr \big( MU - U^+M \big) \Big)^2
  + L_8 \tr \big( MUMU + U^+MU^+M \big) \big)
\nonumber
\\
&&+ L_9 \tr \Big( F^{(+)}_{\mu \nu}D^{\mu}U\,\overline{D}^{\nu}U^+
                 +F^{(-)}_{\mu \nu}\overline{D}^{\mu}U^+\,D^{\nu}U
            \Big)
  - L_{10} \tr \Big( U^+ F^{(+)}_{\mu \nu}U F^{(-)\mu \nu} \Big)
\nonumber
\\
&&+ H_1 \tr \Big( \big( F^{(+)}_{\mu \nu} \big)^2
                 +\big( F^{(-)}_{\mu \nu} \big)^2 \Big)
  + H_2 \tr M^2\,\,,
\label{LeffGL}
\end{eqnarray}
where the  dimensionless structure constants $L_i (i=1,...,10)$ and
$H_{1,2}$ were introduced by Gasser and Leutwyler in ref.
\cite{Gasser-Leutwyler}.
    Here we have introduced the notations
$$
U = \Omega^2 \;\; ;\;\;\; L_{\mu}=D_{\mu}U\,U^+ \;\; ; \;\;\;
F^2_0 = y {{N_c \mu^2}\over{(4 \pi^2)}} \;\;,
$$
$$
\mbox{with}\;\;\;
 y=\Gamma\big(0,\mu^2/\Lambda^2\big) \;\; ; \;\;\;
M = diag(\chi^2_u,\chi^2_d,...,\chi^2_n)\;\; ,\;\;\;
\chi^2_i=m^i_0\mu /(G_1 F^2_0)
= - 2m^i_0\!\!<\!\!\overline{q}q\!\!>\!\! F^{-2}_0 \;;
$$
$<\overline{q} q >$ is the quark condensate;
\begin{eqnarray*}
  F^{(\pm)}_{\mu \nu} =
       \partial_{\mu} A^{(\pm)}_{\nu}
      -\partial_{\nu} A^{(\pm)}_{\mu}
      +[A^{(\pm)}_{\mu},A^{(\pm)}_{\nu} ]
\end{eqnarray*}
are field--strength tensors, and
\begin{eqnarray}
D_{\mu}* = \partial_{\mu}* + (A^{(-)}_{\mu}* - *A^{(+)}_{\mu})\,,
\quad
\overline{D}_{\mu}* = \partial_{\mu}* + (A^{(+)}_{\mu}*-*A^{(-)}_{\mu})
\label{deriv}
\end{eqnarray}
are the covariant derivatives; $A^{(\pm)}_{\mu} =
V_{\mu} \pm A_{\mu}$.
    Moreover, the coefficients $L_i$ and $H_{1,2}$ are given by
$L_1-\frac{1}{2}L_2 = L_4 = L_6 = 0$ and
\begin{eqnarray}
  L_2&=& \frac{N_c}{16 \pi^2}\frac{1}{12}\,,\quad
  L_3 = -\,\frac{N_c}{16 \pi^2}\frac{1}{6}\,,\quad
\nonumber
\\
  L_5&=& \frac{N_c}{16 \pi^2}x(y-1)\,,\quad
  L_7 =  -\,\frac{N_c}{16 \pi^2}\frac{1}{6}\bigg(xy-\frac{1}{12}\bigg)\,,
\nonumber
\\
  L_8  &=& -\frac{N_c}{16 \pi^2}yx^2\,,\quad
  L_9   =   \frac{N_c}{16 \pi^2}\frac{1}{3}\,,\quad
  L_{10}=  -\, \frac{N_c}{16 \pi^2}\frac{1}{6}\,,
\nonumber
\\
  H_1&=&-\,\frac{N_c}{16 \pi^2}\frac{1}{6}\bigg(y-\frac{1}{2}\bigg)\,,\quad
  H_2 = -\,\frac{N_c}{16 \pi^2} 2yx^2\,,
\label{Lcoeff}
\end{eqnarray}
where $x = -\mu F_0^2/(2 <\! \! \overline{q} q \! \! >)$.

    The effective (nonreduced) Lagrangian for the pseudoscalar sector,
taking into account also the emission of the ``structural'' photons
${\cal A}_{\mu}$, can be obtained from (\ref{LeffGL}) when
$V_{\mu}=A_{\mu}=0$ and tensor $F^{(\pm)}_{\mu \nu}$ is replaced by
$ie(\partial_{\mu}{\cal A}_{\nu} - \partial_{\nu}{\cal A}_{\mu})$.
    In the following section we will discuss the reduced nonlinear
Lagrangian for pseudoscalar fields, which arises from generating
functional (\ref{genfunc}) after integrating out the vector and
axial--vector degrees of freedom in the modulus of quark determinant.

%---------------------------------------------------------------------------
%
\section*{3. Strong Lagrangians with reduced vector and axial-vector fields}
%
%---------------------------------------------------------------------------

    To perform the integration over vector and axial--vector fields we
will use the fact that the modulus of quark determinant is invariant
under chiral rotations.
    Then, the pseudoscalar fields can be eliminated from the modulus
of quark determinant in the effective action (\ref{action}) by using
the rotated Dirac operator (\ref{eq4}) for unitary gauge.
    After such transformation the pseudoscalar degrees of freedom
still remain in the mass term of eq.(\ref{action}), quadratic in meson
fields, which are not invariant under chiral rotations.
    Since the masses of the vector and axial--vector mesons are large
compared to the pion mass it is possible to integrate out the rotated
fields $\widetilde{V}_{\mu}\, \mbox{and}\, \widetilde{A}_{\mu}$
(\ref{rotated-fields}) in the effective meson action using the
equations of motion which arise from the mass terms of the effective
action (\ref{action}) in the static limit \cite{statlim-reinhardt}.
    In such an approximation the kinetic terms
$(\widetilde{F}^{(\pm)}_{\mu \nu})^2$\,\, for the rotated fields
$\widetilde{V}_{\mu}\, \mbox{and}\, \widetilde{A}_{\mu}$ as well as
higher order derivative nonanomalous and Wess--Zumino terms are treated
as a perturbation.

    In terms of the rotated fields $\widetilde{V}_{\mu}, \,
\widetilde{A}_{\mu}$ (\ref{rotated-fields}) the quadratic part of the
effective action (\ref{action}) leads to the Lagrangian
\begin{eqnarray}
{\cal L}_0 = \frac{F_0^2}{4} \tr ( MU + h.c. )
            -\bigg (\frac{m^0_V}{g^0_V} \bigg)^2
             \tr [ (\widetilde{V}_{\mu} - v_{\mu} )^2
                  +(\widetilde{A}_{\mu} - a_{\mu} )^2 ]\,,
\label{L0}
\end{eqnarray}
where $(m_V^0 /g^0_V)^2 = 1/(4G_2)$,
with $m^0_V$ and $g^0_V$ being the bare mass and coupling constant of
the vector gauge field, and
\begin{eqnarray*}
v_{\mu} = \frac{1}{2}\Big( \Omega  \partial_\mu \Omega^+
                          +\Omega^+\partial_\mu \Omega  \Big ) ,\quad
a_\mu = \frac{1}{2} \Big ( \Omega  \partial_\mu \Omega^+
                          -\Omega^+\partial_\mu \Omega  \Big ) \, .
\end{eqnarray*}

    The modulus of quark determinant contributes to divergent and
finite parts of the effective meson Lagrangian.
    In terms of the rotated fields, taking into account that for unitary gauge
$\U \rightarrow \Sigma$, the divergent part of the quark
determinant (eq.(\ref{append1}) from the Appendix B) gives
\begin {eqnarray}
{\cal{L}}_{div} &=& \frac{F^2_0}{4 \mu^2}
    \tr \bigg \{ D_\mu \Sigma \, \overline{D}^\mu \Sigma^+
  + \frac{1}{6} \big[ \big( \widetilde{F}^{(+)}_{\mu \nu} \big)^2
                     +\big( \widetilde{F}^{(-)}_{\mu \nu} \big)^2
                \big] \bigg \}
\nonumber
\\
                &=& \frac{F^2_0}{4 \mu^2}
    \tr \bigg \{ [ \widetilde{V}_\mu, m_0 ]^2-\{ \widetilde{A}_\mu,m_0 \}^2
              -4 \mu (2m_0+\mu ) \widetilde{A}^2_\mu
              +\frac{1}{6} \big[ \big( \widetilde{F}^{(+)}_{\mu \nu}\big)^2
              +\big( \widetilde{F}^{(-)}_{\mu \nu} \big)^2 \big] \bigg \}\,,
\label{Ldiv}
\end{eqnarray}
    where the approximation $\Sigma = \mu + m_0$ was used.

    The $p^4$--terms of the finite part of the effective meson
Lagrangians (eq.(\ref{append2}) from the Appendix B) are of the form
\begin{eqnarray}
{\cal {L}}^{(p^4)}_{fin} &=& \frac{N_c}{32 \pi^2 \mu^4} \tr \bigg\{
   \frac{1}{3}\Big[\mu^2 D^2 \Sigma \, \overline{D}^2 \Sigma^+
  -(D^{\mu} \Sigma \, \overline{D}_{\mu} \Sigma^+ )^2 \Big]
  +\frac{1}{6} (D_{\mu} \Sigma \, \overline{D}^{\mu} \Sigma^+ )^2
\nonumber
\\
&&-\mu^2 \Big({\cal M}D_{\mu} \Sigma \, \overline{D}^{\mu} \Sigma^+
  +\overline{\cal M}\, \overline{D}^{\mu} \Sigma^+ \, D_{\mu} \Sigma \Big)
  +\mu^2 \frac{2}{3}\Big( D^{\mu}\Sigma \,\overline{D}^{\nu}
                         \Sigma^+ \,\widetilde{F}^{(-)}_{\mu \nu}
  + \overline{D}^{\mu}\Sigma^+\, D^{\nu}\Sigma
  \,\widetilde{F}^{(+)}_{\mu\nu}
                    \Big )
\nonumber
\\
&&+ \mu^2 \frac{1}{3} \widetilde{F}^{(+)}_{\mu \nu} \Sigma^+
                      \widetilde{F}^{(-)\mu \nu} \Sigma
  - \frac{1}{6} \mu^4 \Big[ (\widetilde{F}^{(-)}_{\mu \nu})^2
                           +(\widetilde{F}^{(+)}_{\mu \nu})^2 \Big]
    \bigg \}
 \nonumber
\\
&=& \frac{N_c}{32 \pi^2} \tr \bigg \{
   -\frac{4}{3}\Big[ 2\widetilde{A}^2_{\mu}\Big( \{\widetilde{A}_{\nu},
                            \{ \widetilde{A}^{\nu}, \overline{m}_0 \}
                            \}
   -[\widetilde{V}_{\nu},[\widetilde{V}^{\nu}, \overline{m}_0 ] ] \Big )
\nonumber
\\
&& +[\widetilde{V}_{\mu},\widetilde{A}^{\mu} ]
    \Big( \{\widetilde{A}_{\nu},[\widetilde{V}^{\nu},\overline{m}_0]\}
         +[\widetilde{V}_{\nu},\{ \widetilde{A}^{\nu},\overline{m}_0\}]
    \Big)
   +[\widetilde{V}_{\mu},\widetilde{A}^{\mu} ]^2 \Big]
\nonumber
\\
&& +\frac{8}{3}\Big( (\widetilde{A}_{\mu}
                       \widetilde{A}_{\nu})^2
   +\{\{\widetilde{A}_{\mu}, \overline{m}_0 \},\widetilde{A}_{\nu}
       \widetilde{A}^{\mu} \widetilde{A}^{\nu} \} \Big)
   + 16 \mu^2 \overline{m}_0 \widetilde{A}^2_{\mu}
\nonumber
\\
&& -\frac{4}{3} \Big[ \Big( 2 \widetilde{A}^{\mu} \widetilde{A}^{\nu}
   +[\widetilde{A}^{\mu}, \{ \widetilde{A}^{\nu},\overline{m}_0 \} ]
                      \Big ) \big( \widetilde{F}^{(-)}_{\mu \nu}
                                  +\widetilde{F}^{(+)}_{\mu \nu} \big)
   -\{ \widetilde{A}^{\mu},[ \widetilde{V}^{\nu}, \overline{m}_0] \}
      (\widetilde{F}^{(+)}_{\mu \nu} - \widetilde{F}^{(-)}_{\mu \nu})
                \Big ]
\nonumber
\\
&& +\frac{1}{3} \Big( \widetilde{F}^{(+)}_{\mu \nu}
                      \widetilde{F}^{(-) \mu \nu}
   + \overline{m}_0\{ \widetilde{F}^{(+)}_{\mu \nu},
                      \widetilde{F}^{(-) \mu \nu} \} \Big )
   - \frac{1}{6} \Big[ (\widetilde{F}^{(-)}_{\mu \nu} )^2
                      +(\widetilde{F}^{(+)}_{\mu \nu} )^2 \Big] \bigg\}
   + O(m^2_0 ),
\label{L4fin}
\end{eqnarray}
where $\overline{m}_0 \equiv m_0/\mu$, and  we used the approximation
${\cal M}=\overline{\cal M}=\Sigma^2-\mu^2\approx 2\mu^2\overline{m}_0$
for matrices ${\cal M}$ and $\overline{\cal M}$ in unitary gauge.

    The kinetic terms $(\widetilde{F}^{(V,A)}_{\mu \nu})$,
arising from the sum of Lagrangians (\ref{Ldiv}) and (\ref{L4fin}),
lead to the standard form after rescaling the rotated nonphysical
vector and axial--vector fields $\widetilde{V}_{\mu}, \widetilde{A}_{\mu}$:
\begin{eqnarray}
  \widetilde{V}_{\mu} = \frac{g^0_V}{(1+\widetilde{\gamma})^{1/2}}
                        \widetilde{V}_{\mu}^{(ph)}\, ,\quad
  \widetilde{A}_{\mu} = \frac{g^0_V}{(1-\widetilde{\gamma})^{1/2}}
                        \widetilde{A}_{\mu}^{(ph)}\, .
\label{rescale}
\end{eqnarray}
    Here
\begin{eqnarray}
   g^0_V = \bigg[ \frac{N_c}{48 \pi^2}
           \bigg( \frac{8\pi^2F^2_0}{N_c{\mu}^2}-1\bigg)\bigg]^{-1/2},
\quad
   \widetilde{\gamma} = \frac{N_c (g^0_V)^2}{48\pi^2}\, ,
\label{param1}
\end{eqnarray}
and $\widetilde{V}_{\mu}^{(ph)}, \widetilde{A}_{\mu}^{(ph)}$ are the
physical fields of vector and axial--vector mesons with masses
\begin{eqnarray}
  m_{\rho}^2 = \frac{(m^0_V)^2}{1+\widetilde{\gamma}}\, ,\quad
  m_{A_1}^2 = \frac{(m^0_V)^2}{1-\widetilde{\gamma}} Z^{-2}_A \, ,
\label{VAmasses}
\end{eqnarray}
where $Z^2_A = 1 - (F_0 g^0_V/ m^0_V)^2$ is the $\pi A_1$-mixing
factor.

    Since in the following we also want to investigate the radiative
processes with ``structural'' photon emission in addition to inner
bremsstrahlung ones, it is necessary to include electromagnetic
interactions in the bosonization procedure.
    Obviously, one then simply has to use the replacements
$\widetilde{V}_{\mu}^{(ph)} \rightarrow
\widetilde{V}_{\mu}^{(ph)} + ie^{(ph)} {\cal A}_{\mu}^{(ph)} Q$ ,
or $\widetilde{V}_{\mu} \rightarrow
\widetilde{V}_{\mu} + ie_0 {\cal A}_{\mu} Q$, where $Q$ is the matrix
of electric quark charges and
$$
{\cal A}_{\mu}^{(ph)}=\frac{g^0_V}{(1+\widetilde{\gamma})^{1/2}}
{\cal A}_{\mu}, \quad
e^{(ph)}=e_0\frac{(1+\widetilde{\gamma})^{1/2}}{g^0_V}
$$
are the physical electromagnetic field and charge respectively.

    The static equations of motion arise from variation the mass terms
of eq.(\ref{L0}) in chiral limit over rotated fields
$\widetilde{V}_{\mu},\,\widetilde{A}_{\mu}$ and lead to the relations
\begin{eqnarray}
\widetilde{V}_{\mu} = v_{\mu}^{(\gamma)}, \quad
\widetilde{A}_{\mu} = Z^2_A \,a^{(\gamma)}_{\mu}
\label{eqmotion1}
\end{eqnarray}
and
\begin{eqnarray}
\widetilde{F}^{(\pm)}_{\mu\nu} &=&
      (Z^4_A-1)[a^{(\gamma)}_{\mu},a^{(\gamma)}_{\nu}]
     + \,ie_0 Q \cal{F}^{(\gamma)}_{\mu \nu}
     + \,\mbox{$ie_0$} (\cal{A}_{\mu} [ \mbox{$Q, v^{(\gamma)}_{\nu}$} ]
     - \cal{A}_{\nu} [ \mbox{$Q, v^{(\gamma)}_{\mu} $} ])
 \nonumber
 \\
&& \pm \mbox{$ie_0$} Z^2_A ({\cal{A}}_{\mu} [ Q,a^{(\gamma)}_{\nu}]
     - {\cal A}_{\nu}[Q, a^{(\gamma)}_{\mu}]).
\label{eqmotion2}
\end{eqnarray}
    Here
\begin{eqnarray*}
v^{(\gamma)}_{\mu} =
    \frac{1}{2}\Big( \Omega \partial^{(\gamma)}_{\mu} \Omega^+
                    +\Omega^+ \partial^{(\gamma)}_{\mu} \Omega \Big) ,
\quad
a^{(\gamma)}_{\mu} =
    \frac{1}{2}\Big( \Omega \partial^{(\gamma)}_{\mu} \Omega^+
                    -\Omega^+ \partial^{(\gamma)}_{\mu} \Omega \Big)
 = -\frac{1}{2}\xi_{R}^{+} L^{(\gamma)}_{\mu} \xi_{R}\,;
\end{eqnarray*}
$\partial^{(\gamma)}_{\mu}*=\partial_{\mu}*+ie_0{\cal{A}}_{\mu}[Q,*]
=\partial_{\mu}*+ie^{(ph)}{\cal{A}}_{\mu}^{(ph)}[Q,*]\,$
is the prolonged derivative describing the emission of the inner
bremsstrahlung photon while the electromagnetic field strength tensor
${\cal F}^{(\gamma)}_{\mu \nu} =
\partial_{\mu}{\cal A}_{\nu} - \partial_{\nu} {\cal A}_{\mu}$
corresponds to the structural photon
$\big( e_0 {\cal F}^{(\gamma)}_{\mu \nu} =
    e^{(ph)} {\cal F}^{(\gamma,ph)}_{\mu \nu} \big)$; and
$L^{(\gamma)}_{\mu} = (\partial^{(\gamma)}_{\mu} U ) U^+$.
    Further, we will omit for simplicity the upper indices
$(\gamma)$ corresponding to the inner bremsstrahlung photon and only
tensors ${\cal F}^{(\gamma)}_{\mu \nu}$ will be kept explicitly.
    We will also omit everywhere the upper indices $(ph)$ assuming
that all photons and electromagnetic charges in further formulae are
physical.

    Applying the equations of motion (\ref{eqmotion1}) to the terms of the
effective actions (\ref{L0},\ref{Ldiv}), quadratic in vector and
axial--vector fields, one reproduces the standard kinetic term for the
pseudoscalar sector:
\begin{eqnarray}
L_{kin} = -\, \frac{F^2_0}{4} \tr \big( L_{\mu} L^{\mu} \big) .
\label{eq10}
\end{eqnarray}
    In the same way the $p^4$--terms of the actions
(\ref{Ldiv},\ref{L4fin}) lead to the reduced Lagrangians for
pseudoscalar mesons of the types
\begin{eqnarray}
{\cal L}^{(p^4,red)} &=&
   \frac{1}{2} \widetilde{L}_2 \tr \Big( [L_{\mu},L_{\nu}]^2 \Big)
 + (3\widetilde{L}_2+\widetilde{L}_3) \tr \Big( (L_{\mu} L^{\mu})^2 \Big)
\nonumber
\\
&& - 2\widetilde{L}_5 \tr \Big( L_{\mu} L^{\mu} \xi_R M \xi^+_R \Big)
   - 2ie {\cal F}^{(\gamma)}_{\mu \nu} \widetilde{L}_9
                   \tr \Big( Q \xi^+_R L^{\mu} L^{\nu} \xi_R \Big )
\nonumber
\\
&& - 2(ie)^2 \widetilde{L}_{10} \tr \Big[ {\cal A}^2_{\mu}
     \Big( Q\xi^+_R L_{\nu} \xi_R \, Q \xi^+_R L^{\nu} \xi_R
     - Q^2 \xi^+_R L^2_{\nu} \xi_R \Big)
\nonumber
\\
&& - {\cal A}_{\mu} {\cal A}_{\nu}
     \Big( Q \xi^+_R L^{\mu} \xi_R \, Q \xi^+_R L^{\nu} \xi_R
     - Q^2 \xi^+_R L^{\mu} L^{\nu} \xi_R \Big) \Big]\, ,
\label{Lp4red}
\end{eqnarray}
\begin{eqnarray}
{\cal L}^{(p^4,red)}_{m_0} &=&
   \frac{N_c}{32 \pi^2} \, \frac{4}{3}
   \tr \Big\{ -4Z^4_A \Big[ (L_{\mu} L^{\mu})^2-2(L_{\mu}L_{\nu})^2
                          + L_{\nu}L_{\mu}L^{\mu}L^{\nu}\Big]
                     \big( \xi_R \overline{m}_0 \xi^+_R \big)
\nonumber
\\
&& + (Z^4_A-1)^2 \Big[ (L_{\mu}L_{\nu})^2
                      - L_{\nu}L_{\mu}L^{\mu}L^{\nu} \Big]
                 \big( \xi_R \overline{m}^0 \xi^+_R \big) \Big\}
 \nonumber
 \\
 && - ie {\cal F}^{(\gamma)}_{\mu \nu} \frac{N_c}{32\pi^2} \frac{1}{3} \,
      \tr \Big\{ Q\Big[ (1+Z^4_A)
              \{ \xi^+_R L^{\mu} L^{\nu}\xi_R,\,\overline{m}_0 \}
    + 4Z^4_A ( \xi^+_R L^{\mu} \xi_R \overline{m}_0 \xi^+_R L^{\nu} \xi_R)
        \Big] \Big\}
 \nonumber
 \\
 && - (ie)^2 \frac{N_c}{32 \pi^2} \frac{1}{3} Z^4_A\,
      \tr \Big\{\overline{m}_0 \Big[
      {\cal A}^2_{\mu} \Big( 2(Q \xi^+_R L_{\nu} \xi_R)^2
      -\xi^+_R L_{\nu} \xi_R \{ \xi^+_R L_{\nu} \xi_R,Q^2 \} \Big)
 \nonumber
 \\
 && - {\cal A}_{\mu} {\cal A}_{\nu}\Big(
      2 Q \xi^+_R L^{\mu} \xi_R\, Q \xi^+_R L^{\nu} \xi_R
    - \xi^+_R L^{\mu} \xi_R
                  \{ \xi^+_R L^{\nu} \xi_R, Q^2 \} \Big) \Big] \Big\}
    + O(\overline{m}^2_0)\,,
\label{Lp4redm0}
\end{eqnarray}
    Here ${\cal L}^{(p^4,red)}$ represents the part, corresponding to
the effective $p^4$--Lagrangian in the Gasser--Leutwyler
representation with the structure coefficients $\widetilde{L}_i$
defined by the relations,
\begin{eqnarray}
\widetilde{L}_2 &=& \frac{N_c}{16 \pi^2}
    \bigg[ \frac{1}{12} Z^8_A
  + \frac{1}{6} (Z^4_A-1) \bigg( (Z^4_A-1)
    \frac{6\pi^2}{N_c} \, \frac{1+\widetilde{\gamma}}{(g^0_V)^2}
  - Z^4_A \bigg) \bigg] ,
\nonumber
\\
\widetilde{L}_3 &=& -\frac{N_c}{16 \pi^2}
    \bigg[ \frac{1}{6} Z^8_A
  + \frac{1}{2} (Z^4_A-1) \bigg( (Z^4_A-1)
    \frac{6\pi^2}{N_c} \, \frac{1+\widetilde{\gamma}}{(g^0_V)^2}
  - Z^4_A \bigg) \bigg] ,
\nonumber
\\
\widetilde{L}_5 &=& \frac{N_c}{16 \pi^2} Z^4_A x(4y-1),
\nonumber
\\
\widetilde{L}_9 &=& \frac{N_c}{16 \pi^2} \bigg[ \frac{1}{3} Z^4_A
  - (Z^4_A-1) \frac{4\pi^2}{N_c} \, \frac{1+\widetilde{\gamma}}{(g^0_V)^2}
                                         \bigg] ,
\nonumber
\\
\widetilde{L}_{10} &=& -Z^4_A \, \frac{1}{4}
                          \, \frac{1+\widetilde{\gamma}}{(g^0_V)^2},
\label{Ltilde}
\end{eqnarray}
where
$$
\frac{1+\widetilde{\gamma}}{(g^0_V)^2} = \frac{F^2_0}{6 {\mu}^2} \, .
$$
    The Lagrangian ${\cal L}^{(p^4,red)}_{m_0}$ describes the additional
corrections arising from the expansion over the quark mass $m_0$.

    To take into account also the $p^6$--terms of the finite part of the
effective meson Lagrangian (eq.(\ref{append3}) from Appendix B) we will
restrict ourselves only by consideration of the terms which
additionally contribute to $m_0$-corrections of type of
eq.(\ref{Lp4redm0}):
\begin{eqnarray}
{\cal {L}}^{(p^6)}_{fin} &=& \frac{N_c}{32 \pi^2 \mu^4} \tr \bigg\{
    \frac{1}{6} \Big[ {\cal M} \Big(
    D_{\mu} D_{\nu} \Sigma \, \overline{D}^{\mu} \overline{D}^{\nu} \Sigma^+
  + D_{\mu} \Sigma \, \overline{D}^2 \overline{D}^{\mu} \Sigma^+
  + D^2 D_{\mu} \Sigma \, \overline{D}^{\mu} \Sigma^+ \Big)
\nonumber
\\
&&+ \overline{\cal M} \Big(
    \overline{D}^{\mu} \overline{D}^{\nu} \Sigma^+ \,D_{\mu} D_{\nu} \Sigma
  + \overline{D}_{\mu} \Sigma^+ \, D^2 D^{\mu} \Sigma
  + \overline{D}^2 \overline{D}^{\mu} \Sigma^+ \, D_{\mu} \Sigma \Big) \Big]
\nonumber
\\
&&- \frac{1}{12{\mu}^2} \Big[ {\cal M} \Big(
    \big( D_{\mu} \Sigma \, \overline{D}^{\mu} \Sigma^+ \big) ^2
  - \big( D_{\mu} \Sigma \, \overline{D}_{\nu} \Sigma^+ \big) ^2
  + D_{\mu} \Sigma \, \overline{D}_{\nu} \Sigma^+ \,
    D^{\nu} \Sigma \, \overline{D}^{\mu} \Sigma^+ \Big)
\nonumber
\\
&&+ \overline{\cal M} \Big(
    \big( \overline{D}_{\mu} \Sigma^+ \, D^{\mu} \Sigma \big) ^2
  - \big( \overline{D}_{\mu} \Sigma^+ \, D_{\nu} \Sigma \big) ^2
  + \overline{D}_{\mu} \Sigma^+ \, D_{\nu} \Sigma
    \overline{D}^{\nu} \Sigma^+ \, D^{\mu} \Sigma \Big) \Big]
\nonumber
\\
&&- \frac{1}{3} \Big[ \widetilde{F}^{(-)}_{\mu \nu} \Big(
    D^{\mu} \Sigma \, \overline{D}^{\nu} \Sigma^+ \, {\cal M}
  + {\cal M} \, D^{\mu} \Sigma \, \overline{D}^{\nu} \Sigma^+
  + D^{\mu} \Sigma \, \overline{\cal M} \, \overline{D}^{\nu} \Sigma^+
    \Big)
\nonumber
\\
&&+ \widetilde{F}^{(+)}_{\mu \nu} \Big(
    \overline{D}^{\mu} \Sigma^+ D^{\nu} \Sigma \, \overline{\cal M}
  + \overline{\cal M} \, \overline{D}^{\mu} \Sigma^+ \, D^{\nu} \Sigma
  + \overline{D}^{\mu} \Sigma^+ \, {\cal M} \, D^{\nu} \Sigma
    \Big) \Big]
\nonumber
\\
&&- \frac{5}{6} {\mu}^2 \Big[
    {\cal M} (\widetilde{F}^{(-)}_{\mu \nu} )^2
  + \overline{\cal M}(\widetilde{F}^{(+)}_{\mu \nu} )^2 \Big]
 + \mbox{(other terms)}\,.
\label{L6fin}
\end{eqnarray}
    All other omitted terms after reducing the vector and axial-vector
fields in eq.(\ref{append3}) will contribute only to the
interaction of six or more pseudoscalar mesons.

    The final expression for the reduced Lagrangian
${\cal L}^{(p^4,red)}_{m_0}$ can be presented  then in the general form
\begin{eqnarray}
{\cal L}^{(p^4,red)}_{m_0} &=&
     \tr \Big[ \Big( \widetilde{Q}_1 \big( L_{\mu} L^{\mu} \big)^2
     + \widetilde{Q}_2 \big( L_{\mu}L_{\nu} \big)^2
     + \widetilde{Q}_3 L_{\nu}L_{\mu}L^{\mu}L^{\nu} \Big)
       \big( \xi_R \overline{m}_0 \xi^+_R \big) \Big]
\nonumber
\\
&& - ie {\cal F}^{(\gamma)}_{\mu \nu} \,
     \tr\Big[ Q\Big(
     \widetilde{Q}_4\big\{\xi^+_RL^{\mu}L^{\nu}\xi_R,\,\overline{m}_0\big\}
     +\widetilde{Q}_5\big(\xi^+_RL^{\mu}\xi_R\overline{m}_0
                          \xi^+_RL^{\nu}\xi_R\big) \Big) \Big]
\nonumber
\\
&& - (ie)^2 \widetilde{Q}_6
     \tr \Big\{ \overline{m}_0 \Big[
              {\cal A}^2_{\mu} \Big( 2( Q \xi^+_R L_{\nu} \xi_R )^2
     - \xi^+_R L_{\nu} \xi_R \{ \xi^+_R L_{\nu} \xi_R,Q^2 \} \Big)
\nonumber
\\
&& - {\cal A}_{\mu} {\cal A}_{\nu}
     \Big(2 Q \xi^+_R L^{\mu} \xi_R\,Q\xi^+_R L^{\nu} \xi_R
   - \xi^+_R L^{\mu} \xi_R
                 \{ \xi^+_R L^{\nu} \xi_R, Q^2 \} \Big) \Big] \Big\}
\label{Lp4m0red}
\end{eqnarray}
with $\widetilde{Q}_i$ being the structure coefficients:
\begin{eqnarray}
\widetilde{Q}_1 &=& \frac{N_c}{16 \pi^2} \frac{1}{6} Z^4_A (5Z^4_A-17),
\nonumber
\\
\widetilde{Q}_2 &=& \frac{N_c}{16 \pi^2} \frac{1}{3}
    \bigg[ Z^8_A + 16Z^4_A + \frac{3}{16} (Z^4_A-1)(17Z^4_A-9) \bigg] ,
\nonumber
\\
\widetilde{Q}_3 &=& -\frac{N_c}{16 \pi^2} \frac{1}{6}
    \bigg[ 3Z^8_A + 16Z^4_A + \frac{3}{8} (Z^4_A-1)(7Z^4_A-9) \bigg] ,
\nonumber
\\
\widetilde{Q}_4 &=& \frac{N_c}{16 \pi^2} \frac{1}{12} (5Z^4_A+7),\quad
\widetilde{Q}_5 = 0, \quad
\widetilde{Q}_6 = \frac{N_c}{16 \pi^2} \frac{3}{8}Z^4_A.
\label{Qparam}
\end{eqnarray}

%----------------------------------------------------------------------
%
\section*{4. Reduced currents}
%
%----------------------------------------------------------------------

    The path--integral bosonization method can be applied to the
weak and electromagnetic--weak currents by using a generating
functional for Green functions of quark currents
introduced in \cite{pich-rafael} and \cite{bosoniz-our}.
    After transition to collective fields in such a generating
functional the latter is determined by the analog of formula
(\ref{logarithm}) where now $i{\bf \widehat{D}}$ is replaced by
\begin{eqnarray}
i{\bf \widehat{D}}(\eta)=
   [ i(\widehat{\partial}+\widehat{A}_R-i\widehat{\eta}_R)
    - (\Phi+m_0-\eta_R) ]P_R
  +[ i(\widehat{\partial}+\widehat{A}_L-i\widehat{\eta}_R)
    - (\Phi^{+}+m_0-\eta_L)]P_L\,.
\label{dirsources}
\end{eqnarray}
    Here
$\widehat{\eta}_{L,R} = \eta_{L,R\mu}^a \gamma^{\mu}\frac{\lambda^a}{2}$
and $\eta_{L,R} = \eta_{L,R}^a \frac{\lambda^a}{2}$ are the external
sources coupling to the quark currents
$\overline{q} P_{L,R} \gamma^{\mu} \frac{\lambda^a}{2}q$ and quark
densities $\overline{q} P_{L,R} \frac{\lambda^a}{2}q$ respectively.
    The quark densities define the contributions of the penguin--type
four--quark operators of the effective nonleptonic weak Lagrangian
\cite{penguin} to the matrix elements of relevant kaon decays.
    The bosonized $(V \mp A)$ and $(S \mp P)$ meson currents,
corresponding to the quark currents
$\overline{q} P_{L,R} \gamma_{\mu} \frac{\lambda^a}{2}q$ and quark
densities $\overline{q} P_{L,R} \frac{\lambda^a}{2}q$, can be obtained
by variating the quark determinant with redefined Dirac operator
(\ref{dirsources}) over the external sources coupling with these
quark bilinears \cite{bosoniz-our}.

    For further discussions it is convenient to present the bosonized
weak and electromagnetic-weak $(V-A)$--current for pseudoscalar sector,
generated by the nonreduced Lagrangian (\ref{LeffGL}) and including
the electromagnetic--weak structural photon emission, in the form:
\begin{eqnarray}
J^{(nred)a}_{L\mu}&=&i\frac{F^2_0}{4}\tr \Big(\lambda^a L_{\mu}\Big)
\nonumber
\\
&& -i\tr \Big\{ \lambda^a \Big[
     \frac{1}{2}R_1U^+\big\{(MU+U^+M),L_{\mu}\big\}U
   + R_2L_{\nu}L_{\mu}L^{\nu}
\nonumber
\\
&& + R_3\big\{ L_{\mu},L_{\nu}L^{\nu}\big\}
   + R_4\partial_{\nu}\big( [L_{\mu},L^{\nu}] \big) \Big]\Big\}
\nonumber
\\
&&+ e {\cal F}^{(\gamma)}_{\mu \nu}R_5\tr\Big(\lambda^a[UQU^+,L^{\nu}]
                                         \Big)\,.
\label{JeffGL}
\end{eqnarray}
Here, the first term is the kinetic current and all other terms
originate from the $p^4$--part of Lagrangian (\ref{LeffGL});
$R_i$ are the structure coefficients, associated with the
corresponding parameters $\widetilde{R}_i$ of the representation
(\ref{VAcurrent}) for the reduced $(V-A)$--currents:
\begin{eqnarray}
R_1=-L_5\,,\quad R_2=2L_2\,,\quad R_3=2L_2+L_3\,,\quad
R_4=-\frac{1}{2}L_9\,,\quad R_5=L_9+L_{10}\,.
\label{RparamGL}
\end{eqnarray}

    The bosonized $(S-P)$ current for pseudoscalar sector, generated
by the Lagrangian (\ref{LeffGL}) and including the structural photons,
has the form:
\begin{eqnarray}
J^{(nred)a}_L &=& \frac{F^2_0}{8\mu}\tr \big(\lambda^a \partial^2U \big)
   + \frac{F^2_0}{4}\mu R \tr \big(\lambda^a U \big)
\nonumber
\\
&& - \frac{1}{\mu} \tr \Big\{ \lambda^a \Big[
     L_2\partial_{\mu}\big( L_{\nu}L^{\mu}L^{\nu} \big)
   + \big( 2L_2+L_3 \big) \partial_{\mu}\big( L_{\nu}L^{\nu}L^{\mu}\big)
\nonumber
\\
&& -\frac{1}{2} L_5 \Big( \partial_{\mu}\big( (MU+U^+M)L^{\mu}U \big)
                         + 2\mu^2 R L_{\mu}L^{\mu} \Big) \Big] \Big\}
\nonumber
\\
&& - \frac{ie}{2\mu} L_9 \tr \Big\{ \lambda^a \Big[
     \partial^\mu\big({\cal F}^{(\gamma)}_{\mu \nu}\,[Q,L^{\nu}U]\big)\Big\}
   - \frac{1}{2\mu} L_{10} \big(ie{\cal F}^{(\gamma)}_{\mu \nu}\big)^2
     \tr \big( \lambda^a QUQ \big)\,.
\label{JSPnred}
\end{eqnarray}
    Here, the first and second terms are generated at $p^2$--level by
the kinetic and mass terms of Lagrangian (\ref{LeffGL}), respectively,
while all other terms originate from its $p^4$--part.

    Combining the method of the chiral bosonization of quark currents
with the static equations of motions it is possible to obtain the
bosonized meson currents for pseudoscalar sector with the reduced
vector and axial--vector degrees of freedom.
    In this way one can reproduce the standard kinetic $(V-A)$ current
for pseudoscalar mesons
\begin{eqnarray*}
J^{(kin)a}_{L\mu} &=& i\frac{F^2_0}{4} \tr
                        \Big( \lambda^a \xi_R^+ L_{\mu} \xi_R \Big)\,,
\end{eqnarray*}
which arises from the terms of effective actions (\ref{L0},\ref{Ldiv}),
quadratic in vector and axial-vector rotated fields, after
redefinition of the rotated fields
$$
\widetilde{V}_{\mu}  \to V_{\mu}-i(\eta_{L\mu}+\eta_{R\mu})\,,\quad
\widetilde{A}_{\mu}  \to A_{\mu}+i(\eta_{L\mu}-\eta_{R\mu})\,,
$$
and variation over $\eta_{L\mu}$ with applying the static equations of motion.

    Applying the same procedure to the $p^4$-- and $p^6$--terms
(\ref{L4fin},\ref{L6fin}) of the effective action we also obtain the
bosonized weak and electromagnetic--weak $(V-A)$ currents for
pseudoscalar sector with the reduced vector and axial--vector degrees
of freedom.
    It is convenient to present these reduced currents in the form:
\begin{eqnarray}
J^{(p^4, red)a}_{L \mu} &=&
   - i\,\widetilde{R}_1 \tr \Big( \lambda^a
      \big\{M, \xi^+_R L_\mu \xi_R\big\} \Big)
\nonumber
\\
&& - i\tr \Big\{ \lambda^a \Big[
      \widetilde{R}_2 \big( \xi^+_R L_{\nu}L_{\mu}L^{\nu} \xi_R \big)
     +\widetilde{R}_3 \big( \xi^+_R \{L_{\mu},L_{\nu}L^{\nu}\}\xi_R \big)
     +\widetilde{R}_4 \partial_{\nu} \big( \xi^+_R [L_{\mu},L^{\nu}] \xi_R
                                     \big)\Big]\Big\}
\nonumber
\\
&& + e{\cal F}^{(\gamma)}_{\mu \nu}\widetilde{R}_5
       \tr \Big( \lambda^a [Q,\xi^+_R L^\nu \xi_R] \Big)\,,
\label{VAcurrent}
\end{eqnarray}
\begin{eqnarray}
J^{(p^4, red)a}_{m_0, L \mu} &=& - i\tr \Big\{ \lambda^a \Big[
     \widetilde{R}_6\{\overline{m}_0,\xi^+_RL_{\nu}L_{\mu}L_{\nu}\xi_R \}
   + \widetilde{R}_7 \Big( \overline{m}_0\xi^+_RL_{\nu}^2L_{\mu}\xi_R
                    +\xi^+_RL_{\mu}L_{\nu}^2\xi_R \overline{m}_0 \Big)
\nonumber
\\
&& + \widetilde{R}_8 \Big( \overline{m}_0\xi^+_RL_{\mu}L_{\nu}^2\xi_R
     +\xi^+_RL_{\nu}^2L_{\mu}\xi_R \overline{m}_0
     +\xi^+_R \{ L_{\mu},L_{\nu}\xi_R \overline{m}_0
                                \xi^+_RL^{\nu} \} \xi_R \Big)
\nonumber
\\
&& + \widetilde{R}_9 \xi^+_R\Big(
      L_{\mu} \xi_R \overline{m}_0 \xi_R^+ L_{\nu}^2
     +L_{\nu}^2 \xi_R \overline{m}_0 \xi^+_R L_{\mu} \Big) \xi_R
\nonumber
\\
&& + \widetilde{R}_{10} \xi^+_R \Big(
       L_{\nu} L_{\mu} \xi_R \overline{m}_0 \xi^+_R L^{\nu}
     + L^{\nu} \xi_R \overline{m}_0 \xi^+_R L_{\mu} L_{\nu} \Big)\xi_R
\nonumber
\\
&& + \widetilde{R}_{11} \partial_{\nu} \Big(
     \xi^+_R \big( L^{\nu}\xi_R\overline{m}_0\xi^+_RL_{\mu}
                -L_{\mu}\xi_R\overline{m}_0\xi^+_RL^{\nu} \big) \xi_R \Big)
   + \widetilde{R}_{12} \big\{\overline{m}_0,
     \partial_{\nu}\big(\xi^+_R[L_{\mu},L^{\nu}]\xi_R\big)\big\}\Big]\Big\}
\nonumber
\\
&& + e{\cal F}^{(\gamma)}_{\mu \nu} \tr \Big\{ \lambda^a \Big[
     \widetilde{R}_{13} \big( Q \xi^+_R L^{\nu} \xi_R \overline{m}_0
                             -\overline{m}_0 \xi^+_R L^{\nu} \xi_R Q \big)
\nonumber
\\
&& + \widetilde{R}_{14} \big( \overline{m}_0 Q \xi^+_R L^{\nu} \xi_R
                             -\xi^+_R L^{\nu} \xi_R Q \overline{m}_0 \big)
   + \widetilde{R}_{15} \big( Q \overline{m}_0 \xi^+_R L^{\nu} \xi_R
                             -\xi^+_R L^{\nu} \xi_R \overline{m}_0 Q \big)
     \Big] \Big\}
\label{VAm0cur}
\end{eqnarray}
    with $\widetilde{R}_i$ being the structure coefficients:
\begin{eqnarray}
\widetilde{R}_1&=&-\,\frac{N_c}{16\pi^2}\,\frac{1}{2}Z^2_Ax(y-1) ,
\nonumber
\\
\widetilde{R}_2&=&\frac{N_c}{16\pi^2}\,\frac{1}{12}\,Z^2_A
    \bigg(Z^4_A + 1 - (Z^4_A-1)
    \frac{12\pi^2}{N_c}\,\frac{1+\widetilde{\gamma}}{(g^0_V)^2}\bigg),
\nonumber
\\
\widetilde{R}_3&=&\frac{1}{2}\widetilde{R}_4=
   -\,\frac{N_c}{16\pi^2}\,\frac{1}{24}\,Z^2_A
    \bigg(1-(Z^4_A-1)
    \frac{12\pi^2}{N_c}\,\frac{1+\widetilde{\gamma}}{(g^0_V)^2}\bigg),
\nonumber
\\
\widetilde{R}_5&=&-\frac{N_c}{16\pi^2}\,\frac{1}{6}\,Z^2_A
    \bigg( 1-
    \frac{12\pi^2}{N_c}\,\frac{1+\widetilde{\gamma}}{(g^0_V)^2}\bigg),
\nonumber
\\
\widetilde{R}_6&=&\frac{N_c}{16\pi^2}\,\frac{1}{96}\,Z^2_A(19Z^4_A-1),
\quad
\widetilde{R}_7 = \frac{5}{6} \widetilde{R}_9
                = -\,\frac{N_c}{16\pi^2}\,\frac{5}{24}Z^6_A,
\nonumber
\\
\widetilde{R}_8&=&-\,\frac{N_c}{16\pi^2}\,\frac{1}{96}\,Z^2_A(19Z^4_A-5),
\quad
\widetilde{R}_{10} = \frac{N_c}{16\pi^2}\,\frac{1}{96}\,Z^2_A(23Z^4_A-1),
\nonumber
\\
\widetilde{R}_{11}&=&\frac{N_c}{16\pi^2}\,\frac{1}{12}\,Z^4_A(Z^2_A-2),
\quad
\nonumber
\\
\widetilde{R}_{12}&=&\frac{N_c}{16\pi^2}\,\frac{1}{24}\,Z^2_A
    \bigg( Z^4_A+4Z^2_A-\frac{9}{2} Z^{-2}_A(Z^4_A-1)\bigg),
\nonumber
\\
\widetilde{R}_{13}&=&\frac{3}{5} \widetilde{R}_{14}
                   = \frac{1}{2} \widetilde{R}_{15}
                   = -\,\frac{N_c}{16 \pi^2}\,\frac{1}{8}Z^2_A.
\label{Rparam}
\end{eqnarray}
    Thus, the reduction of the vector and axial--vector fields does
not change the kinetic term of the bosonized $(V-A)$ current while
the structure of the $p^4$--part of $(V-A)$ current is strongly
modified (compare (\ref{JeffGL}) and (\ref{VAcurrent})).

    Using the bosonization procedure of ref.\cite{bosoniz-our} and the
equations of motion (\ref{eqmotion1}) we obtain also the reduced $(S-P)$
meson currents.
    After redefinition of scalar fields
\begin{eqnarray}
 \Sigma \to \Sigma - 2\eta_R\,, \qquad \Sigma^+ \to \Sigma^+ - 2\eta_L
\label{scalrdf}
\end{eqnarray}
and variation over $\eta_L$ with applying the static equations of
motion the divergent part of the effective action (\ref{Ldiv}) leads
to the scalar current
\begin{eqnarray}
J^{(div,red)a}_L &=& \frac{F^2_0}{8\mu}Z^2_A \tr \Big\{ \lambda^a \Big[
     \xi_R^+L^2_{\mu}\xi_R + 4\mu^2 \overline{m}_0
   + \frac{1}{4} \Big(
     2\big(\xi^+_RL_{\mu}\xi_R\overline{m}_0\xi^+_RL^{\mu}\xi_R \big)
   + \big\{\overline{m}_0,\xi_R^+L^2_{\mu}\xi_R \big\} \Big)\Big]\Big\}
\nonumber
\\
&& + \frac{F^2_0}{4}\mu R Z^{-2}_A \tr\Big(\lambda^a(1+\overline{m}_0)\Big),
\label{Jdiv}
\end{eqnarray}
where $R=<\!\!\overline{q}q\!\!>\!\!/(\mu F^{2}_0)$.

    The $p^4$-- and $p^6$--terms of the finite part of the effective action
(\ref{L4fin},\ref{L6fin}) generate the scalar meson current which can be
present in the form
\begin{eqnarray}
J^{(p^4, red)a}_L &=&
   - \mu \widetilde{G}_1 \tr\Big( \lambda^a \big(\xi_R^+L^2_{\mu}\xi_R
                            \big) \Big)
\nonumber
\\
&& - \frac{1}{\mu} \tr \Big\{ \lambda^a \Big[
     \widetilde{G}_2 \partial^2 \big( \xi_R^+L^2_{\mu}\xi_R \big)
   + \widetilde{G}_3 \Big(
     \big( \partial_\mu \big( \xi_R^+L_{\nu}\xi_R \big) \big)^2
     +\big\{\xi^+_RL_{\mu}\xi_R\,,\partial^2\big(\xi_R^+L^{\mu}\xi_R\big)
      \big\}
                     \Big)
\nonumber
\\
&& + \widetilde{G}_4 \Big(
     \partial_\mu\big(\xi_R^+L_{\nu}\xi_R\big)\,\xi^+_RL^{\mu}L^{\nu}\xi_R
     +\xi^+_RL^{\nu}L^{\mu}\xi_R\,\partial_\mu\big(\xi_R^+L_{\nu}\xi_R\big)
                     \Big)
\nonumber
\\
&& + \widetilde{G}_5 \Big(
     \xi^+_RL^{\mu}\xi_R\,\partial_\mu\big(\xi_R^+L_{\nu}\xi_R\big)\,
     \xi^+_RL^{\nu}\xi_R
   - \xi^+_RL^{\nu}\xi_R\,\partial_\mu\big(\xi_R^+L_{\nu}\xi_R\big)\,
     \xi^+_RL^{\mu}\xi_R
\nonumber
\\
&& + \xi^+_RL^{\mu}L^{\nu}\xi_R\,\partial_\mu\big(\xi_R^+L_{\nu}\xi_R\big)
   - \xi^+_RL^{\nu}L^{\mu}\xi_R\,\partial_\mu\big(\xi_R^+L_{\nu}\xi_R\big)
     \Big)
\nonumber
\\
&& + \widetilde{G}_6 \partial_\mu\big(
     \xi_R^+\big[L_{\nu},[L^{\nu},L^{\mu}]\big]\xi_R\big)
   + \widetilde{G}_7 \big(\xi_R^+L_{\nu}L^2_{\mu}L^{\nu}\xi_R\big)
\nonumber
\\
&& + \widetilde{G}_8 \big(\xi_R^+L^2_{\mu}L^2_{\nu}\xi_R\big)
   + \widetilde{G}_9 \big(\xi_R^+(L_{\mu}L_{\nu})^2\xi_R\big)
     \Big] \Big\}
\nonumber
\\
&& - ie\frac{1}{\mu} \tr \Big\{ \lambda^a \Big[
     \widetilde{G}_{10} \partial^\mu\big({\cal F}^{(\gamma)}_{\mu \nu}\,
     [Q,\xi_R^+L^{\nu}\xi_R]\big)
\nonumber
\\
&& + {\cal F}^{(\gamma)}_{\mu \nu}\,\Big(
     \widetilde{G}_{11}\big(\xi^+_RL^{\mu}\xi_R\,Q\,\xi_R^+L^{\nu}\xi_R\big)
   + \widetilde{G}_{12}\big\{\xi^+_RL^{\mu}L^{\nu}\xi_R,Q\big\}
     \Big) \Big] \Big\}
\nonumber
\\
&& + \big(ie{\cal F}^{(\gamma)}_{\mu \nu}\big)^2\frac{1}{\mu}
   \widetilde{G}_{13} \tr\big( \lambda^a Q^2 \big) + O(\overline{m}_0)
\label{spcurrent}
\end{eqnarray}
where $\widetilde{G}_i$ are the structure coefficients:
\begin{eqnarray}
\widetilde{G}_1&=&6\widetilde{G}_2=6\widetilde{G}_3
                =\frac{3}{2}\widetilde{G}_{11}
                =\frac{N_c}{128\pi^2}Z^4_A\,,\quad
\widetilde{G}_4 = \widetilde{G}_5 =
                  \frac{N_c}{256\pi^2}\,\frac{1}{3}Z^6_A\,,
\nonumber
\\
\widetilde{G}_6&=&\frac{N_c}{256\pi^2}\,\frac{1}{6}\,Z^2_A
                  \big(Z^4_A - 1\big)\,,
\nonumber
\\
\widetilde{G}_7&=&\frac{N_c}{256\pi^2}\,\frac{1}{6}\,
    \Big(2Z^8_A+\big(Z^4_A-1\big)\big(2Z^4_A+1\big)\Big)\,,
\nonumber
\\
\widetilde{G}_8&=&\frac{N_c}{256\pi^2}\,\frac{1}{6}\,
    \Big(2Z^8_A+\big(Z^4_A-1\big)\big(Z^4_A+1\big)\Big)\,,
\nonumber
\\
\widetilde{G}_9&=&\frac{N_c}{256\pi^2}\,\frac{1}{6}\,
    \Big(5Z^8_A-\big(Z^4_A-1\big)\big(3Z^4_A+2\big) \bigg)\,,
\nonumber
\\
\widetilde{G}_{10}&=&\frac{N_c}{256\pi^2}\,\frac{2}{3}Z^2_A\,,\quad
\widetilde{G}_{12} = -\,\frac{N_c}{256\pi^2}\,\frac{1}{3}
                      \big(5Z^4_A-2\big)\,,\quad
\widetilde{G}_{12} = \frac{N_c}{256\pi^2}\,\frac{4}{3}\,.
\label{Gparam}
\end{eqnarray}
    It is important to mention that the $p^6$--terms (\ref{L6fin}) of
the finite part of the effective action contributes only to the
$m_0$--corrections (\ref{VAm0cur}) for $(V-A)$ current while in the
case of $(S-P)$ currents due to the fact that redefinition
(\ref{scalrdf}) leads to the replacement
$$
{\cal M} \to {\cal M} - 2(\eta_R \Sigma^+ + \Sigma \eta_L)\,,\qquad
\overline{\cal M} \to \overline{\cal M} - 2(\Sigma^+ \eta_R +\eta_L \Sigma)\,,
$$
the same $p^6$--terms give contributions also to the part of current
(\ref{spcurrent}) do not  containing quark mass $m_0$.

    It can be easily shown that at the $p^2$--level the reduction of the
vector and axial--vector fields does not change the physical results
for matrix elements of the bosonized gluonic penguin operator.
    In fact, both for the reduced current (\ref{spcurrent}) and for
the first two terms of the nonreduced current (\ref{JSPnred}) the
corresponding $p^2$--contributions to the penguin operator matrix
element, dominating by the interference of the kinetic and mass terms
of the scalar current, can be presented effectively in the same form:
$$
  <T^{(peng)}> \propto -\,\frac{F^4_0}{32}R
                    < ( \partial_\mu U \,\partial^\mu U^+ )_{23} >.
$$
    On the other hand the structure of the $p^4$--part of the pseudoscalar
meson $(S-P)$ current proves to be strongly modified by the reduction
of the vector and axial--vector fields (compare the expressions
(\ref{JSPnred}) and (\ref{spcurrent})).

%--------------------------------------------------------------------
%
\section*{5. Numerical estimations}
%
%--------------------------------------------------------------------

    To discuss some physical consequences for pseudoscalar nonet of
mesons we have to fix initially the numerical values of the various
parameters entering in the reduced Lagrangian and currents.
    The parameters $\chi^2_i$ can be fixed by the spectrum of
pseudoscalar mesons.
    Here we use the values $\chi^2_u = 0.0114$GeV$^2$,
$\chi^2_d = 0.025$GeV$^2$, and $\chi^2_s = 0.47$GeV$^2$.
    To fix other empirical constants of our model we will use the
experimental parameters, listed in Table 1: the masses of $\rho$-- and
$A_1$--mesons, the coupling constants of $\rho \to \pi \pi$ and
$\pi,K \to \mu \nu$ decays, the $\pi\pi$--scattering lengths $a^I_l$,
the pion electromagnetic squared radii $ <r^2_{em}>_{\pi^+}$ and pion
polarizability $\alpha_{\pi^{\pm}}$.
    We also include in our analysis the data on the
$\gamma \gamma \rightarrow \pi^+ \pi^-$ cross section  near to the
threshold (see Fig.1).
    We will use the relations (\ref{param1}), (\ref{VAmasses}),
$g_V=g^0_V(1+\widetilde{\gamma})^{-1/2}$ and
$$
g_{\rho\pi \pi}=g_V \bigg[ 1 + \frac{m^2_{\rho}}{2F^2_0}
                \bigg( \frac{N_c}{48\pi^2}Z_A^4
               -\frac{F^2_0}{24 \mu^2}Z_A^{-4}(1-Z_A^2)^2 \bigg)\bigg]\,.
$$
    Then, using the connection $m^0_s = \widehat{m}_0 \chi^2_s/m^2_{\pi}$,
where $\widehat{m}_0 \equiv (m^0_u+m^0_d)/2$, the splitting of the
constants $F_{\pi,K^{+}}$ can be presented in the form:
\begin{eqnarray*}
F_{\pi,K^+} &=& Z_{\pi,K^+} F_0\bigg[
               1+\big(\overline{m}^0_u + \overline{m}^0_{d,s}\big)
                 \frac {1}{2} Z^2_A
                 \bigg(1-\frac{N_c{\mu}^2}{4\pi^2F_0^2}
                 \bigg) \bigg]
\nonumber
\\
&\approx& F_0 \bigg[ 1+\big(\overline{m}^0_u + \overline{m}^0_{d,s}\big)
                      \frac{1}{2}Z^2_A\big(1+4Z^2_A\big)
                      \bigg(1 - \frac{N_c\mu^2}{4\pi^2F^2_0}
                                \frac{1+Z^2_A}{1+5Z^2_A}
                      \bigg) \bigg] +O\big( \overline{m}^2_0 \big)\,.
\label{Fsplit}
\end{eqnarray*}
     Here
\begin{eqnarray*}
Z_{{\pi},K^+} = \bigg[1+\big(\overline{m}^0_u+\overline{m}^0_{d,s}\big)
                        4Z^4_A\bigg(1-\frac{N_c\mu^2}{16\pi^2F^2_0}\bigg)
                \bigg]^{1/2}
\end{eqnarray*}
is the factor arising from the renormalization of the pseudoscalar
meson fields
\begin{eqnarray*}
\Phi \to \Phi + \Big\{ 4Z^4_A\bigg(1-\frac{N_c\mu^2}{16\pi^2F^2_0}
                               \bigg) \overline{m}_0\,,\Phi \Big\}
\end{eqnarray*}
which leads the bilinear kinetic part, including terms up to order
$m^0_q$, to the standard form of the kinetic Lagrangian:
\begin{eqnarray*}
{\cal L}^{bil}_{kin}=\frac{1}{2} \tr \bigg\{ \bigg[
      1-8Z^4_A\bigg(1-\frac{N_c\mu^2}{16\pi^2F^2_0}\bigg)
              \overline{m}_0\bigg]
              \partial_{\mu}\Phi\partial^{\mu}\Phi\bigg\}
      \to \frac{1}{2} \tr \big(\partial_{\mu}\Phi\partial^{\mu}\Phi\big)\,.
\end{eqnarray*}

    The $\pi\pi$--scattering lengths are defined by the structure
coefficients $\widetilde{L}_2$ and $\widetilde{L}_3$.
    For $\pi\pi$--scattering lengths $a^I_l$ (indices $I$ and $l$
refer here to the isotopic spin and orbital momentum, respectively)
in one--loop approximation we obtained \cite{bebunper}
\begin{eqnarray*}
a^0_0 &=& \frac{\pi}{2}\alpha_0(9-5\delta)
         +\frac{\pi}{2}\alpha^2_0\Big[5A+3B+2D+3C-6(\delta^2+4b+3)\Big]\,,
\\
a^2_0 &=&-\frac{\pi}{2}\alpha_02\delta
         +\frac{\pi}{2}\alpha^2_02\Big[A+D-3(\delta^2+b+3)\Big]\,,
\\
a^1_1 &=& \frac{\pi}{2}\alpha_0
         +\frac{\pi}{2}\alpha^2_0\frac{1}{3}\bigg[
          B+6\delta+a-3+\frac{1}{3}(\delta^2-b-3)\bigg]\,,
\\
a^0_2 &=& \frac{\pi}{2}\alpha^2_0\bigg[\frac{1}{15}(C+4D)
         -\frac{2}{5}\bigg(5+\frac{3\delta-2a+6}{9}
         -\frac{\delta^2+4b+3}{15}\bigg) \bigg]\,,
\\
a^2_2 &=& \frac{\pi}{2}\alpha^2_0\bigg[\frac{1}{15}(C+D)
         -\frac{1}{5}\bigg(4+\frac{6\delta-a+3}{9}
         -\frac{2}{45}(\delta^2+b+3)\bigg) \bigg] \,.
\end{eqnarray*}
    Here $\alpha_0 = \frac{1}{3}\big(m_\pi/(2\pi F_0)\big)^2\,;\,\,
\delta = \frac{3}{2}(1-\beta)$, with $\beta$ being the parameter of chiral
symmetry breaking which takes here the value $\beta=1/2\,;$
$a = 21(1-\delta)$ and $b=(11\delta^2-15\delta+3)$.
    The parameters
$$
  A = A^B + A^{loop}\,,\quad B = B^B + B^{loop}\,,\quad
  C = C^B + C^{loop}\,,\quad D = D^B + D^{loop}
$$
include in themselves the Born contributions
$$
  A^B =-144\pi^2(\widetilde{L}_2-\widetilde{L}_3)\,,\quad
  B^B =-576\pi^2\widetilde{L}_3\,,\quad
  C^B = 576\pi^2(\widetilde{L}_2+\widetilde{L}_3)\,,\quad
  D^B = 576\pi^2\widetilde{L}_2
$$
and the pion--loop contributions calculated, using the superpropagator
method \cite{SPvolkov}, in ref.\cite{pipi}:
$$
  A^{loop} =-1.5\,,\quad B^{loop} = 3\,,\quad
  C^{loop} = 5.5\,,\quad D^{loop} = 11.
$$

%    The $\eta' \to \eta2\pi$ decay amplitudes in Born approximations
%are
%\begin{eqnarray*}
%T^B_{\eta'\to 2\pi^0\eta} &=&T^B_{\eta' \to \pi^+\pi^-\eta}
%\\
%&=&\frac{\sqrt{2}}{3F^2_0}\bigg(\cos2\phi-\frac{\sin2\phi}{2\sqrt{2}}
%                          \bigg)
%   \bigg\{ m^2_{\pi}
%          -4(3L_2+L_3)\frac{1}{F^2_0}\bigg[
%           2\bigg(3s^2_0+m^2_{\pi}(m^2_{\eta'}+m^2_{\eta})
%\\
%&&-(m^2_{\eta}+m^2_{\pi})(m^2_{\eta'}+m^2_{\pi})\bigg)
%  -\frac{1}{2}(s_1-s_2)^2 - \frac{3}{2}(s_0-s_3)^2\bigg]\bigg\}\,,
%\end{eqnarray*}
%with Dalitz variables $s_i$ given by
%\begin{eqnarray*}
%  s_{1,2} = (p_{\eta'}-p_{\pi_{1,2}})^2\,,\quad
%  s_3 = (p_{\eta'}-p_{\eta})^2\,,\quad
%  s_0 = \frac{1}{3}\big(m^2_{\eta'}+m^2_{\eta}+2m^2_{\pi}\big)\,.
%\end{eqnarray*}
%    The pion--loop contribution to the decay amplitude is strongly
%suppressed due to an additional factor $\alpha_0$, and it will
%therefore be neglected in the following numerical estimates.
%    The fit of the experimental data on the $\pi \pi$--scattering
%lengths and the $\eta' \to \eta 2\pi$ branching ratios, presented in
%Table 1, give
%$$     L_2^{exp} = (1.41 \pm 0.05 )\cdot 10^{-3}\,, \quad
%       L_3^{exp} = -(3.50 \pm 0.16 )\cdot 10^{-3}\,.
%$$

%--------------------------------------------------------------------------
%
%            Table 1
%
%--------------------------------------------------------------------------
\begin{table}[hbt]
\begin{center}
\caption{ Physical input parameters used for the fixing of the
          empirical constants of the model}
\vspace{0.5cm}
\begin{tabular}{|c|c|c|}
\hline
Input parameters & Experiment & Theory
\\ \hline
$m_{\rho}$       & $770 MeV$&  $772 MeV$
\\
$m_{A_1}$        &$1260 MeV$& $1160 MeV$
\\
$g_{\rho\pi\pi}$ &  $6.3$   &  $6.8$
\\
$F_{\pi}$        &  $93 MeV$& $93.9 MeV$
\\
$F_{K}$          & $119 MeV$&$118.5 MeV$
\\ \hline
$a^0_0 \cdot m_{\pi}$ & $0.23 \pm 0.05$ \cite{pipi-belkov} &$0.20$
\\
$a^2_0 \cdot m_{\pi}$ & $-0.05 \pm 0.03$ \cite{pipi-belkov} &$-0.04$
\\
$a^1_1 \cdot m^3_{\pi}$ & $0.036 \pm 0.010$ \cite{pipi-belkov} &$0.038$
\\
$a^0_2 \cdot m^5_{\pi}$ & $(17 \pm 3)\cdot 10^{-4}$ \cite{pipi-nagles} &
                                                    $17 \cdot 10^{-4}$
\\
$a^2_2 \cdot m^5_{\pi}$ & $(1.3 \pm 3)\cdot 10^{-4}$ \cite{pipi-nagles} &
                                                    $2 \cdot 10^{-4}$
\\ \hline
$<r^2_{em}>_{\pi^+}$ & $(0.439\pm 0.030)fm^2$ \cite{dally} & $0.53 fm^2$
\\
$\alpha_{\pi^{\pm}}$ & $(6.8\pm 1.4)\cdot 10^{-4} fm^3$ \cite{antipov}&
$8.0 \cdot 10^{-4} fm^3$
\\ \hline
\end{tabular}
\end{center}
\end{table}
%--------------------------------------------------------------------------

    The pion electromagnetic squared radius is defined as the coefficient
of the $q^2$--expansion of the electromagnetic form factor
$f^{em}_{\pi}(q^2)$:
\begin{eqnarray*}
    <\pi (p_2)|V^{em}_{\mu}|\pi (p_1)>=f^{em}_{\pi}(q^2)(p_1-p_2)_{\mu},
\\
    f^{em}_{\pi}(q^2) = 1 + \frac{1}{6} <r^2_{em}>_{\pi} q^2 + ...
\end{eqnarray*}
    Being restricted only by pion loops, one gets in the
$SP$--regularization the corresponding contribution to the
electromagnetic squared radius \cite{EMRADvolkov,BOOKvolkov}:
\begin{eqnarray*}
<r^2_{em}>^{(loop)}_{\pi^+} =  - \frac{1}{(4 \pi F_0)^2} \bigg[
   3 {\cal{C}} + \ln \Big( \frac{m_ \pi}{2 \pi F_0} \Big)^2 - 1 \bigg]
                            = 0.062fm^2 \, ,
\end{eqnarray*}
   where ${\cal{C}}=0.577$ is the Euler constant.
     Because the main contribution to this value arises from the
logarithmic term, the kaon loop contribution, containing the small
logarithm $ \ln \Big(m_K / (2 \pi F_0) \Big) ^2$, can be neglected.
     At the Born level, the contribution to the pion electromagnetic
squared radius originates from the $\widetilde{L}_9$--term of the
reduced Lagrangian (\ref{Lp4red}):
\begin{eqnarray*}
       <r^2_{em}>^{(Born)}_{\pi^+} = \frac{12}{F_0^2} \widetilde{L}_9
\end{eqnarray*}

   The pion polarizability can be determined through the Compton--scattering
amplitude:
$$
<\pi_1(p_1) \pi_2(p_2)|S|\gamma_{{\lambda}_1}(q_1)\gamma_{{\lambda}_2}(q_2)> =
      T_1 (p_1 p_2 \; | \; q_1 q_2) + T_2(p_1 p_2 | q_1 q_2) \, ,
$$
$$
T^{(\pm)}_1 = 2e^2 \varepsilon^{\nu}_{\lambda_1} \varepsilon^{\nu}_{\lambda_2}
              \bigg( g^{\mu \nu}
                    - \frac{p_1^{\mu}p_2^{\nu}}{p_1 q_1}
                    - \frac{p_1^{\nu}p_2^{\mu}}{p_2 q_1} \bigg),
\quad
  T^{(0)}_1 = 0;
$$
$$
        T_2 = \varepsilon^{\mu}_{\lambda_1} \varepsilon^{\nu}_{\lambda_2}
              \big((q_1 q_2) g_{\mu \nu} - q_{1 \nu} q_{2\mu} \big)
              \beta(q_1 q_2),
$$
where $\beta(q_1 q_2)$ is the so--called dynamical polarizability function.
    Defining the polarizability of a meson as the coefficient of the effective
interaction with the external electromagnetic field
$$
        V_{int} = -\alpha_{\pi} (E^2 - H^2)/2
$$
    one obtains
$$
        \alpha_{\pi} = \frac{\beta_{\pi}(q_1 q_2)}{8 \pi m_{\pi}}\;
                   \bigg|_{(q_1 q_2) = 0} .
$$

    The pion--loops give the finite contributions without $UV$--divergences
[6,7,11]:
\begin{eqnarray*}
   \beta^{(loop)}_{\pi^{\pm}} = \frac{e^2}{8 \pi^2 F^2_0}
          \bigg( 1 - \frac{4\delta - 3}{3\bar s_{\pi}} \bigg)
          f(\bar s_{\pi})\;,
\qquad
   \beta^{(loop)}_{\pi^0} = \frac{e^2}{4 \pi^2 F^2_0}
          \bigg( 1 -\frac{\delta}{3 \bar s_{\pi}} \bigg)f(\bar s_{\pi})\;,
\end{eqnarray*}
   where $\bar s_{\pi}=(q_1 q_2)/(2m^2_{\pi})$ ,
         $f(\zeta) = \zeta^{-1} J^2(\zeta) - 1\;,$ and
\begin{eqnarray*}
   J(\zeta) = \left \{ \begin{array}{ll}
              \arctan (\zeta^{-1} - 1)^{-1/2}, & 0 < \zeta < 1;\\
              \frac{1}{2} \bigg(\ln \frac{1 + \sqrt{1-\zeta^{-1}}}
                                         {1 - \sqrt{1+\zeta^{-1}}}
                                - i\pi \bigg), & \zeta > 1;\\
              \frac{1}{2} \ln \frac{1 + \sqrt{1 - \zeta^{-1}}}
                                   {-1 +\sqrt {1 - \zeta^{-1}}} , & \zeta < 0.
\end{array} \right.
\end{eqnarray*}
    The meson--loop contributions to the pion polarizabilities are
\begin{eqnarray*}
  \alpha^{(loop)}_{\pi^{\pm}} = 0,
\qquad
  \alpha^{(loop)}_{\pi^0} = -\,\frac{e^2}{384 \pi^3 F^2_0 m^2_\pi}
                          = -5.43 \cdot 10^{-5} fm^3\,.
\end{eqnarray*}
    At the Born level, the $\widetilde{L}_9$-- and
$\widetilde{L}_{10}$--terms of the reduced Lagrangian (\ref{Lp4red}) give:
\begin{eqnarray*}
  \beta^{(Born)}_{\pi^{\pm}} =
  \frac{8e^2}{F^2_0} \big(\widetilde{L}_9+\widetilde{L}_{10}\big)\,,
\qquad
  \beta^{(Born)}_{\pi^0} = 0\,.
\end{eqnarray*}
%    Using the fixed value of the coefficient $L_9^{(exp)}$ (\ref{L9})
%and the experimental result of pion polarizability measurement in the
%radiation $\pi$ meson scattering in nuclei Coulomb field \cite{antipov}
%\begin{eqnarray*}
%     \alpha^{(exp)}_{\pi^\pm} = (6.8 \pm 1.4) \times 10^{-4} fm^3
%\end{eqnarray*}
%one can also fix the values of the combination $(L_9+L_{10})$ and
%coefficient $L_{10}$:
%\begin{eqnarray}
%    (L_9+L_{10})^{exp} = (3.5 \pm 0.7)\cdot 10^{-3},\qquad
%     L_{10}^{exp} = -(3.2 \pm 0.8 )\cdot 10^{-3}.
%\label{L10}
%\end{eqnarray}

    In our analysis the constants $F_0$, $\widehat{m}_0$, $\mu$ and
$m^0_V$ are treated as the independent empirical parameters and their
values are fixed as
\begin{eqnarray}
F_0 = 91.7 \mbox{MeV}\,,\quad \widehat{m}_0 = 2.2 \mbox{MeV}\,,\quad
\mu = 186 \mbox{MeV}\,, \quad  m^0_V = 840 \mbox{MeV}\,.
\label{paramfix}
\end{eqnarray}
    The corresponding calculated values of the input parameters are
also presented in Table 1.
The results for the $\gamma \gamma \rightarrow \pi^+ \pi^-$ cross
sections are shown in Fig.1. All other constants can be calculated
using the values (\ref{paramfix}):
$$
  g^0_V = 5.4\,,\quad \widetilde{\gamma} = 0.185\,,\quad
  Z^2_A = 0.653\,,\quad
  <\!\!\overline{q}q\!\!> = - (330 \mbox{MeV})^3\,,\quad
  m^0_s = 53 \mbox{MeV}\,.
$$
    The values for the current quark masses seem to be by a factor of
$2\div 3$ too small and the quark condensate, respectively, by the
same factor too large as compared with the corresponding value from
the usual phenomenological analysis, based on nonreduced Lagrangian
and currents.
    A similar shift of the current masses and the condensate was
already observed, for example, in ref.\cite{alkofer-reinhardt} after
taking into account the vector--scalar and axial--vector--pseudoscalar
mixing in the analysis of the collective mesons mass spectrum within
the extended NJL model.

    Using the values of the parameters $Z^2_A\,, \widetilde{\gamma}\,\,
\mbox{and}\,\,(g^0_V)^2$ which were fixed above, one can compare
numerically the structural parameters $\widetilde{L}_i$ (\ref{Ltilde})
of the reduced effective Lagrangian (\ref{Lp4red}) with the
corresponding parameters $L_i$ of the nonreduced Lagrangian
(\ref{LeffGL}):
\begin{eqnarray}
&\widetilde{L}_2 = 1.20 L_2 = 1.90 \cdot 10^{-3}\,,\quad
 \widetilde{L}_3 = 1.71 L_3 = -5.41 \cdot 10^{-3}\,,\quad
 \widetilde{L}_5 = 1.99 \cdot 10^{-3}\,,
\nonumber
\\
&\widetilde{L}_9 = 1.35 L_9 = 8.53 \cdot 10^{-3}\,,\quad
 \widetilde{L}_{10} = 1.36 L_{10} = -4.33 \cdot 10^{-3}\,,
\label{Lmodif}
\end{eqnarray}
    After substituting the values of $Z^2_A\,, \widetilde{\gamma}$ and
$(g^0_V)^2$ into eqs.(\ref{Rparam}) one can also compare numerically the
structure parameters $\widetilde{R}_i$ and $R_i$:
\begin{eqnarray}
&\widetilde{R}_1 = -0.285 \cdot 10^{-3}\,,\quad
 \widetilde{R}_2 = 0.76 R_2 = 2.42 \cdot 10^{-3}\,,\quad
 \widetilde{R}_3 = -0.992 \cdot 10^{-3} \;\;\; (R_3=0)\,,
\nonumber
\\
&\widetilde{R}_4 = 2\widetilde{R}_3 = 0.62 R_4 = -1.98 \cdot 10^{-3}\,,\quad
 \widetilde{R}_5 = 0.39 R_5 = 1.23 \cdot 10^{-3}\,.
\label{Rmodif}
\end{eqnarray}

    The electromagnetic--weak part of the non reduced current (\ref{JeffGL})
corresponding to the structural constant $R_5$ (respectively, the
$\widetilde{R}_5$--term of the reduced current (\ref{VAcurrent}))
describes the axial--vector form factor $F_A$ of the radiative decay
$\pi \rightarrow l \nu \gamma$.
    The form factors of this decay are defined by the parameterization
of the amplitude
\begin{eqnarray*}
 T_\mu \big(K,\pi \rightarrow l \nu \gamma \big) = \sqrt{2} e \Big[
 F_V \varepsilon_{\mu \nu \alpha \beta} k^{\nu} q^{\alpha}
 \varepsilon^\beta + i F_A \Big( \varepsilon_\mu \big(k  q \big) -
 q_\mu (k  \varepsilon ) \Big) \Big],
\end{eqnarray*}
  where $k$ is the 4--momentum of the decaying meson, $q$ and $
\varepsilon$ are the 4--momentum and polarization 4--vector of the
photon, and the vector form factor $F_V$ is determined by the
anomalous Wess--Zumino electromagnetic--weak current, originating from
the anomalous part of the effective meson action, which is related to
the phase of the quark determinant.
    The ratio of the axial--vector and vector form factors is
determined by the relation
$$
  \frac{F_A}{F_V} = 32\pi^2 R_5\,.
$$
    The theoretical value of the ratio $F_A/F_V=32\pi^2(L_9+L_{10})=1$
arising from non reduced current (\ref{JeffGL}) with structure
constants $L_{9,10}$ (\ref{Lcoeff}) is in disagreement with the
experimental results on this ratio:
\begin{eqnarray*}
 \bigg(\frac{F_A}{F_V}\bigg)^{(exp)} = \left\{ \begin{array}{ll}
		         0.25 \pm 0.12 \quad \cite{PIENUG-LAMPF}, \\
		         0.41 \pm 0.23 \quad \cite{PIENUG-bolotov}. \\
\end{array} \right.
\end{eqnarray*}
    At the same time the $\widetilde{R}_5$ gives the value
\begin{eqnarray*}
 \frac{F_A}{F_V} = -Z^2_A \bigg( 1-\frac{12\pi^2}{N_c}\,
                         \frac{1+\widetilde{\gamma}}{(g^0_V)^2}\bigg)
                 = 0.39
\end{eqnarray*}
in agreement with the experimental data.

    This result reproduces in nonlinear realization of chiral
symmetry the well known result of ref.\cite{SJPN-volkov}, where the
role of $\pi A_1$--mixing in $\pi \rightarrow l \nu \gamma$ decay
was considered in linear parameterization.
    Thus, after reducing the vector and axial--vector degrees of
freedom it proves to be possible to remove the inselfconsistency
in the description of the ratio $F_A/F_V$ and pion polarizability
which arises seemingly in the pseudoscalar sector of the non reduced
effective Lagrangian (\ref{LeffGL}) with the $L_{9,10}$--terms
\footnote{This point was already noted shortly in our resent paper
\cite{raddec-our}.}
(see the detailed discussion of this inselfconsistency, for example,
in ref. \cite{L9L10-donoghue,electroweak-holstein,semilept-bijnens}).
    The same problem was also considered in ref.\cite{bellucci}, where
the values of the structure constants combination $(L_9+L_{10})$ and
pion polarizability $\alpha_{\pi^\pm}$ determined from the fit of
$\gamma \gamma \rightarrow \pi^+ \pi^-$ cross section data were
discussed.
    Fig.1 shows that within the experimental errors the MARK--II data
\cite{MARK-II} are consistent with the experimental result for pion
polarizability obtained from radiative $\pi$ scattering in nuclear
Coulomb fields \cite{antipov}.
   We have taken into account one--loop corrections, while this was not
done in ref.\cite{bellucci}.
   The description of the $\gamma \gamma \rightarrow \pi^+ \pi^-$  cross
section data above $m_{\pi \pi} = 500$MeV can be improved if one takes
into account the unitary corrections in a more complete way
\cite{pennington}.

%--------------------------------------------------------------------
%
\section*{Conclusion}
%
%--------------------------------------------------------------------

   In this paper we considered the modification of the bosonized
Lagrangian and of the currents for the pseudoscalar sector obtained
after integrating out the vector and axial--vector collective fields
in the generating functional of the NJL model.
   It has been shown, that the reduction of the meson resonances does
not affect the kinetic terms of the strong Lagrangian and the
bosonized $(V-A)$ current as well as the $(S-P)$ current, generated by
the divergent part of quark determinant.
   On the other hand, the reduction of the vector and axial--vector
fields leads to an essential modification of those part of the
pseudoscalar strong Lagrangian and of the currents, which originate
from $O(p^4)$ terms of the quark determinant.
   The reduced Lagrangians and currents allow us to take into account
in a simple way all effects arising from resonance exchange
contributions and $\pi A_1$--mixing when calculating the amplitudes of
various processes with pseudoscalar mesons in the initial and final
states.
\\
\\
   The authors are grateful to G.Ecker for useful discussions and
helpful comments.
   One of the authors (A.A.Bel'kov) thanks the Institute of Elementary
Particle Physics, Humboldt-University, Berlin and DESY--Institute for
High Energy Physics, Zeuthen for their hospitality.
He is also grateful to the DFG for support of investigations (Project Eb
139/1--1).
   Another of the authors (A.V.Lanyov) is grateful for the
hospitality extended to him at the DESY--Institute for High Energy
Physics, Zeuthen.

\newpage
%--------------------------------------------------------------------
%
\section*{Appendix A. Heat-kernel computation of quark determinant}
%
%--------------------------------------------------------------------

    The logarithm of the modulus of the quark determinant is defined
in ``proper-time'' regularization as
\begin{eqnarray}
\log |\det i{\bf \widehat{D}}| \,  =
 -\, {1\over 2} \Trp \log({\bf \widehat{D}}^{+}{\bf \widehat{D}}) =
 -\,{1\over 2} \int^{\infty }_{1/\Lambda^2} d \tau \,\frac{1}{\tau}
  \Trp \mbox{exp} \big(-{\bf\widehat{D}}^{+}{\bf\widehat{D}}\tau\big)
\label{proper-time}
\end{eqnarray}
    with $\Lambda$ being the intrinsic regularization parameter.
    The main idea of the heat-kernel method is to evaluate
$$
<x\mid \exp (-{\bf \widehat{D}}^{+}{\bf \widehat{D}\tau})\mid y>
$$
around its nonperturbated part
$$
<x\mid \exp (-(\hbox{\square }+\mu^{2})\tau)\mid y> =
{1\over (4\pi \tau)^{2}} e^{-\mu^{2}\tau+(x-y)^{2}/(4\tau)}
$$
in powers of proper-time $\tau$ with the so-called Seeley--deWitt
coefficients $h_{k}(x,y)$
$$
<x\mid \exp (-{\bf \widehat{D}}^{+}{\bf \widehat{D}\tau})\mid y> =
{1\over (4\pi \tau)^{2}} e^{-\mu^{2}\tau+(x-y)^{2}/(4\tau)}
\sum^{}_{k} h_{k}(x,y)\cdot \tau^{k}.
$$
    After integration over $\tau$ in (\ref{proper-time}) one gets the
expression (\ref{logarithm}) for $\log \mid \det i \bf \widehat{D}\mid$.
    Using the definition of the gamma function $\Gamma (\alpha,x)$,
one can separate the divergent and finite parts of the quark determinant
$$
{1\over 2}\log (\det {\bf \widehat{D}}^{+} {\bf \widehat{D}}) =
          B_{\hbox{pol}}+ B_{\hbox{log}}+ B_{\hbox{fin}},
$$
    where
$$
B_{\mbox{pol}} = {1\over 2} {e^{-x} \over (4\pi )^{2}}
                 \left[- {\mu^{4}\over 2x^{2}} \Trp h_{0}
                +{1\over x}({\mu^{4}\over 2} \Trp h_{0}
                -\mu^{2}\Trp h_{1} )\right]
$$
has a pole at $x=\mu^2 / \Lambda^2=0$,
$$
B_{\hbox{log}} = -{1\over 2} {1\over (4\pi )^{2}}\Gamma (0,x)
                 \left[ {1\over 2}\mu^{4}\Trp h_{0}-\mu^{2}\Trp h_{1}
                 +\Trp h_{2} \right]
$$
is logarithmic divergent, and the finite part has the form
$$
B_{\hbox{fin}} = - {1\over 2} {1\over (4\pi )^{2}}
                   \sum_{k=3}^{\infty} \mu^{4-2k}\Gamma (k-2,x)\Trp
                   h_{k}.
$$

   The very lengthy calculations of the Seeley-deWitt coefficients
$h_k$ can be performed only by computer support.
   The calculation of the heat-coefficients is a recursive
process which can be done by Computer Algebra Systems such as FORM and
REDUCE very conveniently.
   In ref.\cite{heat-our} we have calculated the full coefficients
up to the order $k=6$.
   After voluminous computations one gets the complex expressions
for heat-coefficients $h_1, \ldots h_4$:
\begin{eqnarray*}
h_{0}(x) &=& 1 ,\\
h_{1}(x) &=& -a ,\\
\Tr [h_{2}(x)] &=& \Tr \left\{
   {1\over 12}(\Gamma_{\mu\nu})^{2} + {1\over 2}a^{2}\right\},
\\
\Tr [h_{3}(x)] &=& - {1\over 12} \Tr \left\{
  {2a^{3}
 - S_\mu S^\mu
 + a(\Gamma_{\mu\nu})^{2}
 - {2\over 45}(K_{\alpha\beta\gamma })^{2}
 - {1\over 9}(K^\alpha {}_{\alpha\beta})^{2}}\right\},
\\
\Tr [h_{4}(x)] &=& \Tr \Bigg\{
   {1\over 24}a^{4}
 + {1\over 12}{a^{2} S^\mu {}_\mu
 + a S_\mu S^\mu }
 + {1\over 720}{7(S^\mu {}_\mu )^{2}
 -(S_{\mu\nu})^{2}}
\\&&
 + {1\over 30}a^{2}(\Gamma_{\mu\nu})^{2}
 + {1\over 120}(a\Gamma_{\mu\nu})^{2}
 + {1\over 180}a(K^\alpha {}_{\alpha\mu})^{2}
 + {1\over 75}a\Gamma_{\mu\nu}K_\beta {}^{\beta\mu\nu}
 + {7\over 900}\Gamma_{\mu\nu}S^\mu K_\alpha {}^{\alpha\nu}
\\&&
 + {1\over 50}aK_\beta {}^{\beta\mu\nu}\Gamma_{\mu\nu}
 - {1\over 300}\Gamma_{\mu\nu}K_\alpha {}^{\alpha\mu}S^\nu
 + {1\over 3600}K^\alpha {}_{\alpha\mu}
      \left(S_\beta {}^{\beta\mu} + S_\beta{}^{\mu\beta} \right)
 + {1\over 72}S_\mu {}^\mu (\Gamma_{\alpha\beta})^{2}
\\&&
 + {1\over 180}S^{\mu\nu}\{\Gamma_{\mu\alpha},\Gamma_\nu {}^\alpha \}
 + {1\over 40}a \left(\Gamma_{\mu\nu}S^{\mu\nu}
                      + {11\over 9}S_{\mu\nu}\Gamma^{\mu\nu} \right)
 + {1\over 144}a \left[K^\mu {}_{\mu\nu},S^\nu \right]
\\&&
 + \left(  {2\over 135}aK_{\beta\mu\nu}
         + {11\over 900}\Gamma_{\mu\nu}S_\beta
         + {1\over 100}S_\beta \Gamma_{\mu\nu}
         + {1\over 4725}\left[\Gamma_{\mu\nu},K^\alpha_{\alpha\beta}\right]
  \right)
  (K^{\beta\mu\nu} -K^{\mu\nu\beta})
\\&&
 + {1\over 1260} \Gamma_{\mu\nu} K_\alpha {}^{\alpha\mu} K_\beta{}^{\beta\nu}
 - {1\over 12600} \left(
         29\Gamma^\beta {}_\alpha \Gamma^{\mu\alpha}
       + 27\Gamma^{\mu\alpha}\Gamma^\beta {}_\alpha
    \right)
    \left( K^\nu_{\mu\beta\nu}
          + K^\nu_{\mu\nu\beta} \right)
\\&&
 + \Gamma_{\alpha\beta}\Gamma_{\mu\nu}
       \left({83\over 25200}K^{\mu\nu\alpha\beta}
           + {4\over 1575}K^{\alpha\beta\mu\nu}
           - {127\over 5040}K^{\alpha\mu\nu\beta}
           - {1\over 600}K^{\mu\alpha\beta\nu}
       \right)
\\&&
 + {13\over 12600}\Gamma_{\mu\beta} \Gamma^\beta{}_\nu \Gamma^\nu{}_\alpha
\Gamma^{\alpha\mu}
 + {47\over 16800}(\Gamma_{\mu\nu})^{2}(\Gamma_{\alpha\beta})^{2}
 + {17\over 25200}(\Gamma_{\mu\nu}\Gamma_{\alpha\beta})^{2}
\\&&
 + {4\over 1575}(\Gamma_{\mu\alpha}\Gamma^\alpha{}_\nu)^{2}
 + {19\over 25200}K^\alpha{}_{\alpha\mu\nu}K^\mu{}_\beta{}^{\beta\nu}
 - {1\over 12600}(K^\alpha{}_{\mu\nu\alpha})^{2}
 + {1\over 1575}(K_\mu{}^\alpha{}{}_{\alpha\nu})^{2}
\\&&
 + {1\over 6300} K_\mu{}^\alpha{}_{\alpha\nu} K^{\beta\mu\nu}{}_\beta
 + {1\over 5600}(K^\alpha{}_{\alpha\mu\nu})^{2}
 - {1\over 5040}{K^\alpha{}_{\alpha\mu\nu}K_\beta{}^{\mu\nu\beta}
 + K_{\mu\nu\alpha\beta}K^{\alpha\mu\nu\beta}}
\\&&
 - {1\over 1800} K_{\mu\alpha}{}^\alpha{}_\nu K^{\mu\beta}{}_\beta{}^\nu
 - {1\over 25200} K_{\mu\nu\al\be}
      \left[
       3 \left( K^{\mu\nu\al\be} + K^{\nu\al\be\mu} \right)
     + 2 \left( K^{\mu\al\be\nu} + K^{\al\nu\mu\be} \right)
      \right]
\Bigg\}
\end{eqnarray*}
    Here,
$$ K_{\mu\nu} = [d_\mu ,d_\nu ] =
     \Gamma_{\mu\nu} ,\hspace{0.3cm} K_{\lambda \mu\nu} =
     [d_{\lambda },K_{\mu\nu}] ,\hspace{0.3cm} K_{\kappa \lambda \mu\nu} =
     [d_{\kappa },K_{\lambda \mu\nu}]\hspace{0.3cm} \hbox{, etc.} $$
$$ S_\mu  = [d_\mu ,a] ,\hspace{0.3cm} S_{\mu\nu} =
        [d_\mu ,S_\nu ] , \hspace{0.3cm}S_{\lambda \mu\nu} =
        [d_{\lambda },S_{\mu\nu}] \hspace{0.3cm} \hbox{, etc.} $$
are commutators of the operators $d_\mu$ and $a$ which are defined by
the relations
$$
d_\mu = \partial_\mu + \Gamma_\mu, \quad
\Gamma_\mu = V_\mu + A_\mu \gamma^{5}, \quad
a(x) = i\widehat{\nabla} H + H^+H + \frac{1}{4}[\gamma^\mu , \gamma^\nu ]
        \Gamma_{\mu\nu} - \mu^2.
$$
    We  used  the  following notations:
$$
  H=P_{R}\Phi + P_{L}\Phi^{+} = S+i\gamma_{5}P\,,\quad
  \Gamma_{\mu\nu}=[d_\mu ,d_\nu ] =
  \partial_\mu \Gamma_\nu -\partial_\nu \Gamma_\mu
  +[\Gamma_\mu ,\Gamma_\nu ]=F^{V}_{\mu\nu}+\gamma^{5}F^{A}_{\mu\nu}
$$
with $F^{V,A}_{\mu\nu}$ as field strength tensors,
$$F^{V}_{\mu\nu}=
\partial_\mu V_\nu -\partial_\nu V_\mu +[V_\mu ,V_\nu ]+[A_\mu ,A_\nu ],$$
$$F^{A}_{\mu\nu}=
\partial_\mu A_\nu -\partial_\nu A_\mu +[V_\mu ,A_\nu ]+[A_\mu ,V_\nu ],$$
and
$$\nabla_\mu H=\partial_\mu H+[V_\mu ,H]-\gamma^{5}\{A_\mu ,H\}$$
as the covariant derivative.

   In the same way the next orders of heat expansion coefficients
$h_i$ can be obtained using the developed computational technique
based on the usage of computer algebra \cite{heat-our}.
   For simplicity we present below expressions only for minimal
parts of heat-coefficients, i.e., only for the parts which do
not vanish in the pseudoscalar region of the theory when $V_\mu=A_\mu=0$:
\begin{eqnarray*}
\Tr [h_{5}(x)^{min}] &=&
 - \Tr \Bigg\{
  {1\over 120} a^2 (  a^3
                    + S_\mu S^\mu)
 + {1\over 180} a^3 S_\mu{}^\mu
 + 2(a S_\mu)^2
\\&&
 + {1\over 6300}\left[  10 a S_\mu (S^{\mu\nu}{}_\nu
                       + S_\nu{}^{\nu\mu} )
                       - 2a (S_{\mu\nu})^{2}
                       + 17a (S^\mu_\mu )^{2}
                       + S^{\mu\nu}{}_\nu S_\mu
                       + 3 a S_\mu{}^{\mu\nu} S_\nu
                \right]
\\&&
 + {11\over 1008}S_\mu S^\mu S_\nu{}^\nu
 + {19\over 2800}S_\mu S_\nu S^{\mu\nu}
 + {2\over 225}S_\mu S_\nu S^{\nu\mu}
\\&&
 + {1\over 25200}\left[ {3(S^\mu{}_{\mu\nu})^{2}
 -2(S_{\mu\nu\alpha})^{2}
 -23(S^{\mu\nu}{}_\nu )^{2}
 + 7S^\mu{}_{\mu\nu} S^{\nu\alpha}{}_\alpha }\right] \Bigg\}
\\
\Tr [h_{6}(x)^{min }] &=& \Tr \Bigg\{ {1\over 720}a^{2}(a^{4}
 + 4S_\mu a S^\mu )
 + {1\over 420}a^{3}S_\mu S^\mu
\\&&
 + {1\over 20160}a^{2}\left[
   20a^{2}S_\mu{}^\mu
 + 5S_\mu (S^{\mu\nu}{}_\nu
 + S_\nu{}^{\nu\mu} )
 + S^\mu{}_{\mu\nu}S^\nu
 -(S_{\mu\nu})^{2}
 + 11(S_\mu{}^\mu )^{2}
 + 9S^{\mu\nu}{}_\nu S_\mu
                       \right]
\\&&
 + {1\over 25200}a\left[{S_\mu a(73S^{\mu\nu}{}_\nu
 + 37S_\nu{}^{\nu\mu} )
 + 5S^{\mu\nu}(S_\mu S_\nu
 + 4S_\nu S_\mu )}\right]
 + {1\over 2016} a S^\mu{}_\mu S^\nu S_\nu
\\&&
 + {1\over 9450} a S_\mu \left(  37S^{\mu\nu}
                               + 23S^{\nu\mu}
                         \right) S_\nu
 + {1\over 9072} a S_\mu
        \left(  S^{\mu\nu}{}_\nu{}^\alpha{}_\alpha
              + S_\nu{}^{\nu\mu\alpha}{}_\alpha
              + S_\nu{}^\nu{}_\alpha{}^{\alpha\mu}
        \right)
\\&&
 + a S^\mu \left(  {23\over 4800} S_\mu S_\nu{}^\nu
                 + {937\over 302400} S^\nu S_{\mu\nu}
                 + {23\over 10800} S^\nu S_{\nu\mu}
            \right)
 + {1\over 252} a S_\mu S^\nu{}_\nu S^\mu
\\&&
 + {1\over 352800}aS^\mu{}_{\mu\nu} \left(  52S_\alpha{}^{\alpha\nu}
                                          + 53S^{\nu\alpha}{}_\alpha
                                    \right)
 + aS^{\mu\nu}{}_\nu \left(  {1\over 3360}S^\alpha{}_{\alpha\mu}
                            -{1\over 11340}S_{\mu\alpha}{}^\alpha
		     \right)
\\&&
 + {17\over 226800}a(S_{\mu\nu\alpha})^{2}
 + {1\over 317520}a \left(  23S^\mu{}_\mu{}^{\nu\alpha}{}_\alpha
                          + 5S^\mu{}_{\mu\alpha}{}^{\alpha\nu}
                          + 77S^{\nu\mu}{}_\mu{}^\alpha{}_\alpha
		    \right) S_\nu
\\&&
 - {1\over 30240}\left[{53(S_\mu S^\mu )^{2}
 + (S_\mu S_\nu )^{2}}\right]
 + {1\over 352800}S_\mu S_\nu \left(  157S^\mu{}_\alpha{}^{\al\nu}
                                    + 298S^{\mu\nu\alpha}{}_\alpha
			      \right)
\\&&
 + {1\over 70560}S_\mu S_\nu \left(  31S_\alpha{}^{\alpha\nu\mu}
				   + 58S^{\nu\mu\alpha}{}_\alpha
				   + 47S^{\nu\alpha}{}_\alpha{}^\mu
			     \right)
 + {1\over 720}S^\mu S_\mu S^\nu{}_\nu{}^\alpha{}_\alpha
\\&&
 + {5\over 14112}S_\mu S_\nu S_\alpha{}^{\alpha\mu\nu}
 + {1\over 105840}S_\mu \left[  {S^{\mu\nu} \left(  37S^\alpha{}_{\alpha\nu}
                                                  + 70S_{\nu\alpha}{}^\alpha
					    \right)
                              + 35S^{\nu\mu}S^\alpha{}_{\nu\alpha}}
			\right]
\\&&
 + {1\over 21168}S_\mu \left[  {S^{\nu\mu}S^\alpha{}_{\alpha\nu}
			     + S_{\nu\alpha}(5S^{\nu\mu\alpha}
			     + 2S^{\nu\alpha\mu})}
 			\right]
 + {1\over 2880}S_\mu \left(  S^{\mu\nu}{}_\nu
                            + S_\nu{}^{\nu\mu}
			\right) S_\alpha{}^\alpha
\\&&
 + S_\mu S^\nu{}_\nu \left(  {83\over 141120}S^{\mu\alpha}{}_\alpha
                           + {1\over 9408}S_\alpha{}^{\alpha\mu}
		     \right)
 + {1\over 30240}S_\mu \left(  17S_{\nu\alpha}{}^\alpha
                             + 13S^\alpha{}_{\alpha\nu}
			\right) S^{\nu\mu}
\\&&
 + {1\over 7560}S_\mu \left[  2 \left(  S^{\mu\nu\alpha}
			              + S^{\nu\mu\alpha}
			              + S^{\nu\alpha\mu}
                                \right) S_{\nu\alpha}
			    + \left( S_{\nu\alpha}{}^\alpha
			            + 2S^\alpha{}_{\alpha\nu}
                              \right) S^{\mu\nu}
		       \right]
 + {1\over 2160}S_\mu S_{\nu\alpha}S^{\mu\nu\alpha}
\\&&
 - {1\over 635040} \left(  701(S_\mu{}^\mu )^{3}
                         + 583S^{\mu\nu}S_{\mu\alpha}S^\alpha_\nu
		   \right)
 - {689\over 316386}S^\alpha{}_\alpha (S^{\mu\nu})^{2}
\\&&
 - {2\over 2835}S^\mu{}_\mu S^{\nu\alpha} S_{\alpha\nu}
 - {1\over 952560}S^{\mu\nu} \left(  619 S^\alpha{}_\nu S_{\mu\alpha}
                                   + 190 S_{\mu\alpha} S^\alpha{}_\nu
			     \right)
\\&&
 + {1\over 151200} \left[  11 (S^\mu{}_\mu{}^\nu{}_\nu)^2
                         - 2 (S_{\mu\nu}{}^\alpha{}_\alpha)^2
                   \right]
 + {1\over 176400} \left(  (S^\mu{}_{\mu\nu\alpha})^{2}
                         + S^\mu{}_{\mu\nu\alpha} S^{\nu\beta}{}_\beta{}^\alpha
		   \right)
\\&&
 - {1\over 226800}(S_{\mu\nu\alpha\beta})^{2}
 - {103\over 12700800}(S_\mu{}^\alpha{}_{\alpha\nu})^{2}
+ {1\over 66150}S^\mu{}_{\mu\nu\alpha} S^{\nu\alpha\beta}{}_\beta
\\&&
 - {1\over 52920} S^\mu{}_\mu{}^\nu{}_\nu S^{\alpha\beta}{}_{\beta\alpha}
 - {13\over 604800} S^{\mu\nu\alpha}{}_\alpha  S_{\mu\beta}{}^\beta{}_\nu
      \Bigg\} .
\end{eqnarray*}
   To obtain these expressions for the heat coefficients, we have
made extensive use of the cyclic properties of the trace and
of the Jacobi identity.

%--------------------------------------------------------------------
%
\section*{Appendix B. Bosonized effective Lagrangians}
%
%--------------------------------------------------------------------

   The effective meson Lagrangians in terms of collective
fields can be obtained from the quark determinant after calculating in
$\trp h_i(x)$ the trace over Dirac indices.
    The ``divergent'' part of the effective meson Lagrangian is
defined by the coefficients $h_0, h_1$ and $h_2$ of the expansion
(\ref{logarithm}):
\begin{eqnarray}
{\cal L}_{div} &=&
    \frac{N_c}{32 \pi^2} \tr \bigg\{
    \Gamma \bigg( 0,\frac{\mu^2}{\Lambda^2} \bigg) \bigg[
    D^{\mu}\U\,\overline{D}_{\mu}\U^+ - {\cal M}^2
    +\frac{1}{6}\,\bigg((F^{(-)}_{\mu \nu})^2 + (F^{(+)}_{\mu \nu})^2
                  \bigg) \bigg]
\nonumber
\\
&&  + 2 \bigg[ \Lambda^2 e^{-\mu^2 / \Lambda^2}
    - \mu^2 \Gamma \bigg( 0,\frac{\mu^2}{\Lambda^2} \bigg) \bigg]
      {\cal M} \bigg\},
\label{append1}
\end{eqnarray}
    where $D^{\mu}$ and $\overline{D}_{\mu}$ are covariant derivatives
defined by eq.(\ref{deriv}), and ${\cal M} = \U\U^+ - \mu^2$.

   Applying the properties of the covariant derivatives
\begin{eqnarray*}
D_{\mu}(O_1O_2)&=&(D_{\mu}O_1)O_2 + O_1(\overline{D}^{'}_{\mu}O_2)
                =(D_{\mu}^{'}O_1)O_2 + O_1(D_{\mu}O_2),
\\
\overline{D}_{\mu}(O_1O_2)&=&(\overline{D}_{\mu} O_1) O_2
                            +O_1(D^{'}_{\mu} O_2)
                           = (\overline{D}^{'}_{\mu} O_1) O_2
                            +O_1(\overline{D}_\mu O_2  ) ;
\end{eqnarray*}
$$
[ D_{\mu}, D_{\nu} ] O =
  F^{(-)}_{\mu \nu}O - OF^{(+)}_{\mu \nu}, \qquad
  [\overline{D}_{\mu}, \overline{D}_{\nu}]O =
  F^{(+)}_{\mu \nu} O - O F^{(-)}_{\mu \nu}
$$
with
$$D^{'}_{\mu}*= \partial_{\mu}*+[A^{(-)}_{\mu},*], \quad
\overline{D}^{'}_{\mu}* = \partial_{\mu}*+[A^{(+)}_{\mu},*]\,,
$$
    and assuming the approximation
$\Gamma ( k,\mu^2 / \Lambda^2 ) \approx \Gamma(k)$ (valid for $k \geq 1$, and
$\mu^2 / \Lambda^2 \ll 1)$ one can present the $p^4$-terms
of the finite part of the effective meson Lagrangian in the form
\begin{eqnarray}
{\cal L}^{(p^4)}_{fin} &=&
    \frac{N_c}{32 \pi^2 \mu^4} \tr \bigg\{
    \frac{1}{3}\,\Big[ \mu^2 D^2 \U\,\overline{D}^2\U
                      -\big(D^{\mu}\U\,\overline{D}_{\mu}\U^+\big)^2\Big]
  + \frac{1}{6} \big(D_{\mu}\U\,\overline{D}_{\nu}\U^+\big)^2
\nonumber
\\
&&- \mu^2 \big( {\cal M}D_{\mu}\U\,\overline{D}^{\mu}\U^+
         +\overline{\cal M}\,\overline{D}_{\mu} \U^+ D_{\mu}\U \big)
\nonumber
\\
&&+ \frac{2}{3}\mu^2\,\Big(
    D^{\mu}\U\,\overline{D}^{\nu}\U^+\,F^{(-)}_{\mu \nu}
    +\overline{D}^{\mu}\U^+\,D^{\nu}\U\,F^{(+)}_{\mu \nu} \Big)
  + \frac{1}{3}\mu^2 F^{(+)}_{\mu \nu}\U^+F^{(-)\mu \nu} \U
\nonumber
\\
&&- \frac{1}{6}\mu^4\,\Big[(F^{(-)}_{\mu \nu})^2 + (F^{(+)}_{\mu \nu})^2\Big]
    \bigg\},
\label{append2}
\end{eqnarray}
    where $\overline{\cal M}=\U^+\U-\mu^2$.

    In an analogous way, the $p^6$-terms of the finite part of the effective
meson Lagrangian can be presented as
\begin{eqnarray}
{\cal L}^{(p^6)}_{fin} &=&
\frac{N_c}{32\pi^2\mu^6} \tr \bigg\{
    \frac{1}{30}\mu^2
    D^2D_{\alpha}\U\,\overline{D}^2\overline{D}^{\alpha}\U^+
\nonumber
\\
&&+ \frac{1}{6}\mu^2\Big[ {\cal M}
    \Big( D_{\mu}D_{\nu}\U\,\overline{D}^{\mu}\overline{D}^{\nu}\U^+
         + D^{\mu}\U\,\overline{D}^2\overline{D}_{\mu}\U^+
         + D^2D_{\mu}\U\,\overline{D}^{\mu}\U^+ \Big)
\nonumber
\\
&&+ \overline{\cal M}
    \Big( \overline{D}^{\mu}\overline{D}^{\nu}\U^+\,D_{\mu}D_{\nu}\U
         + \overline{D}^{\mu}\U^+\,D^2D_{\mu}\U
         + \overline{D}^2\overline{D}_{\mu}\U^+\,D^{\mu}\U\Big)\Big]
\nonumber
\\
&&+ \frac{1}{3}\mu^2
    \Big( \overline{D}^{\mu}\U^+\,\overline{\cal M}D_{\mu}\U\,{\cal M}
         + {\cal M}^2D_{\mu}\U\,\overline{D}^{\mu}\U^+
         + \overline{\cal M}^2\overline{D}_{\mu}\U^+\,D^{\mu}\U \Big)
\nonumber
\\
&&- \frac{1}{45}
    \Big[\big(D_{\mu}\U\,\overline{D}_{\nu}\overline{D}^{\mu}\U^+\big)^2
    + D_{\mu}\U\,\overline{D}_{\nu}\overline{D}_{\alpha}\U^+
    \big( D^{\alpha}\U\,\overline{D}^{\nu}\overline{D}^{\mu}\U^+
         - D^{\mu}\U\,\overline{D}^{\nu}\overline{D}^{\alpha}\U^+ \big)
\nonumber
\\
&&+ \big( \overline{D}_{\mu} \U^+\,D_{\nu}D^{\mu}\U\big)^2
    +D_{\mu}\U^+\,D_{\nu}D_{\alpha}\U
    \big( \overline{D}^{\alpha}\U^+\,D^{\nu}D^{\mu}\U
         - \overline{D}^{\mu}\U^+\,D^{\nu}D^{\alpha}\U \big) \Big]
\nonumber
\\
&&- \frac{1}{60}\Big[
    D_{\mu}\U\,\overline{D}^{\mu}\U^+
    D_{\nu}D_{\alpha}\U\,\overline{D}^{\nu}\overline{D}^{\alpha}\U^+
  + D_{\mu}\U\,\overline{D}_{\nu}\U^+\big(
    D_{\alpha}D^{\nu}\U\,\overline{D}^{\alpha}\overline{D}^{\mu}\U^+
    - D_{\alpha}D^{\mu}\U\,\overline{D}^{\alpha}D^{\nu}\U^+\big)
\nonumber
\\
&&+ \overline{D}_{\mu}\U^+D^{\mu}\U\,
    \overline{D}_{\nu}\overline{D}_{\alpha}\U^+D^{\nu}D^{\alpha}\U
  + \overline{D}_{\mu}\U^+D_{\nu}\U\big(
    \overline{D}_{\alpha}\overline{D}^{\nu}\U^+D^{\alpha}D^{\mu}\U
  - \overline{D}_{\alpha}\overline{D}^{\mu}\U^+D^{\alpha}D^{\nu}\U
    \big) \Big]
\nonumber
\\
&&- \frac{1}{90}\Big[ D_{\mu}\U\,D^{\mu}\U^+D_{\nu}\U\,
                      \overline{D}^2\overline{D}^{\nu}\U^+
    + D_{\mu}\U\,\overline{D}_{\nu}\U^+ \big(
      D^{\nu}\U\,\overline{D}^2\overline{D}^{\mu}\U^+
    - D^{\mu}\U\,D^2D^{\nu}\U^+ \big)
\nonumber
\\
&&+ D_{\mu}\U^+D^{\mu}\U\,\overline{D}_{\nu}\U^+D^2D^{\nu}\U
    + \overline{D}_{\mu}\U^+D_{\nu}\U \big(
      \overline{D}^{\nu}\U^+D^2D^{\mu}\U
    - \overline{D}^{\mu}\U^+D^2D^{\nu}\U \big) \Big]
\nonumber
\\
&&- \frac{1}{12}\Big[ {\cal M} \Big(
      \big( D_{\mu}\U\,\overline{D}^{\mu}\U^+ \big)^2
    - \big( D_{\mu}\U\,\overline{D}_{\nu}\U^+ \big)^2
    + D_{\mu}\U\,\overline{D}_{\nu}\U^+D^{\nu}\U\,\overline{D}^{\mu}\U^+
      \Big)
\nonumber
\\
&&+ \overline{\cal M}\Big( \big( \overline{D}_{\mu}\U^+D^{\mu}\U \big)^2
    - \big(\overline{D}_{\mu}\U^+\,D_{\nu}\U \big)^2
    + \overline{D}_{\mu}\U^+D_{\nu}\U\,\overline{D}^{\nu}\U^+D^{\mu}\U
      \Big) \Big]
\nonumber
\\
&&+ \frac{1}{180\mu^2} \Big[ \overline{D}_{\mu}\U^+D_{\nu}\U\,
    \overline{D}_{\alpha}\U^+D^{\mu}\U\,\overline{D}^{\nu}\U^+D^{\alpha}\U
    - 2\big( \overline{D}_{\mu}\U^+D^{\mu}\U \big)^3
    + 6\overline{D}_{\mu}\U^+D^{\mu}\U\big(
                         \overline{D}_{\nu}\U^+D_{\alpha}\U \big)^2
\nonumber
\\
&&- 3\big( \overline{D}_{\mu}\U^+D^{\mu}\U\,D_{\nu}\U^+D_{\alpha}\U\,
           \overline{D}^{\alpha}\U^+D^{\nu}\U
    + \overline{D}_{\mu}\U^+D_{\nu}\U\,\overline{D}_{\alpha}\U^+
      D^{\mu}\U\,\overline{D}^{\alpha}\U^+D^{\nu}\U \big) \Big]
\nonumber
\\
&&- \frac{1}{6}\mu^6 \Big[ \big( D^{'}_{\mu}{\cal M}\big)^2
              +\big(\overline{D}^{'}_{\mu} \overline{\cal M}\big)^2\Big]
\nonumber
\\
&&+ \frac{1}{6}\mu^2 \Big[ F^{(-)}_{\mu \nu} \Big(
    \frac{1}{5}D^{\mu}D^{\alpha}\U\,
    \overline{D}^{\nu}\overline{D}_{\alpha}\U^+
    - D^2D^{\mu}\U\,\overline{D}^{\nu}\U^+
    - D^{\mu}\U\,\overline{D}^2\overline{D}^{\nu}\U^+
    - D^{\alpha}D^{\mu}\U\,\overline{D}_{\alpha}\overline{D}^{\nu}\U^+
\nonumber
\\
&&- \frac{13}{60} \big(
    D_{\alpha}\U\,\overline{D}^{\mu}\overline{D}^{\nu}
    \overline{D}^{\alpha}\U^+
  - D^{\mu}D^{\nu}D^{\alpha}\U\,\overline{D}_{\alpha}\U^+\big) \Big)
\nonumber
\\
&&+ F^{(+)}_{\mu \nu} \Big(
    \frac{1}{5}\overline{D}^{\mu}\overline{D}^{\alpha}\U^+
               D^{\nu}\,D_{\alpha}\U
    - \overline{D}^2\overline{D}^{\mu}\U^+D^{\nu}\U
    - \overline{D}^{\mu}\U^+D^2D^{\nu}\U
    - \overline{D}^{\alpha}\overline{D}^{\mu}\U^+D_{\alpha}D^{\nu}\U
\nonumber
\\
&&- \frac{13}{60} \big(
      \overline{D}_{\alpha}\U^+D^{\mu}D^{\nu}D^{\alpha}\U
    - \overline{D}^{\mu}\overline{D}^{\nu}\overline{D}^{\alpha}\U^+
                D_{\alpha}\U \big) \Big) \Big]
\nonumber
\\
&&+ \frac{11}{180}\mu^2F^{(+)}_{\mu \nu}\overline{D}_{\alpha} \U^+\,
                  F^{(-)\mu \nu}D^{\alpha}\U
  + \frac{19}{360}\mu^2 \Big[
    \big(F^{(-)}_{\mu \nu}\big)^2D_{\alpha}\U\,\overline{D}^{\alpha}\U^+
    +\big(F^{(+)}_{\mu \nu})^2\overline{D}_{\alpha}\U^+\,D^{\alpha}\U
    \Big]
\nonumber
\\
&&- \frac{1}{3}\mu^2 \Big[ F^{(-)}_{\mu \nu}\,\big(
                    D^{\mu}\U\,\overline{D}^{\nu}\U^+{\cal M}
                  + {\cal M} D^{\mu}\U\,\overline{D}^{\nu}\U^+
                  + D^{\mu}\U\,\overline{\cal M}\,\overline{D}^{\nu}\U^+
                  \big)
\nonumber
\\
&&+ F^{(+)}_{\mu \nu} \big(
\overline{D}^{\mu}\U^+D^{\nu}\U\,\overline{\cal M}
      + \overline{\cal M}\,\overline{D}^{\mu}\U^+D^{\nu}\U
      + \overline{D}^{\mu}\U^+{\cal M}D^{\nu}\U\big) \Big]
\nonumber
\\
&&+ \frac{1}{3}\mu^2 \Big[ F^{(-)}_{\mu \alpha}F^{(-)\alpha}_{\nu}
    \big(
          D^{\mu}\U\,\overline{D}^{\nu}\U^+
         -D^{\nu}\U\,\overline{D}^{\mu}\U^+ \big)
  + F^{(+)}_{\mu \alpha} F^{(+)\alpha}_{\nu}
    \big(
          \overline{D}^{\mu}\U^+D^{\nu}\U
         -\overline{D}^{\nu}\U^+D^{\mu}\U \big)
\nonumber
\\
&&- F^{(-)}_{\mu \nu}D_{\alpha}\U\,F^{(+) \nu \alpha}\overline{D}^{\mu}\U^+
  - F^{(-)}_{\nu \alpha}D^{\alpha}\U\,F^{(+) \mu \nu}\overline{D}_{\mu}\U^+
    \Big]
\nonumber
\\
&&+ \frac{1}{12} \Big[ F^{(-)}_{\mu \nu}\Big(
    \big\{ D_{\alpha}\U\,\overline{D}^{\alpha}\U^+,
           D^{\mu}\U\,\overline{D}^{\nu}\U^+ \big\}
  + D_{\alpha}\U\,\overline{D}^{\mu}\U^+\big(
           D^{\nu}\U\,\overline{D}^{\alpha}\U^+
          -D^{\alpha}\U\,\overline{D}^{\nu}\U^+ \big)
\nonumber
\\
&&+ D^{\mu}\U\,\overline{D}_{\alpha}\U^+\big(
           D^{\alpha}\U\,\overline{D}^{\nu}\U^+
          -D^{\nu}\U\,\overline{D}^{\alpha}\U^+\big) \Big)
\nonumber
\\
&&+ F^{(+)}_{\mu \nu}\Big( \big\{
    \overline{D}_{\alpha}\U^+D^{\alpha}\U,
    \overline{D}^{\mu}\U^+D^{\nu}\U\big\}
  + \overline{D}_{\alpha}\U^+D^{\mu}\U\big(
    \overline{D}^{\nu}\U^+D^{\alpha}\U
  - \overline{D}^{\alpha}\U^+D^{\nu}\U\big)
\nonumber
\\
&&+ \overline{D}^{\mu}\U^+D_{\alpha}\U\big(
    \overline{D}^{\alpha}\U^+D^{\nu}\U
  - \overline{D}_{\nu}\U^+D^{\alpha}\U\big) \Big) \Big]
\nonumber
\\
&&- \frac{5}{6}\mu^4 \Big[ {\cal M} \big(F^{(-)}_{\mu \nu}\big)^2
              +\overline{\cal M}\big( F^{(+)}_{\mu \nu}\big)^2\Big]
  + \frac{1}{20}\mu^4 \Big[
    F^{(-)}_{\mu \nu}\big\{ {\cal M}\,,D^{'\mu} D^{'\nu}{\cal M} \big\}
  + F^{(+)}_{\mu \nu}\big\{ \overline{\cal M}\,,\overline{D}^{'\mu}
                            \overline{D}^{'\nu }\overline{\cal M}\big\}\Big]
\nonumber
\\
&&- \frac{1}{3}\mu^2
    \Big( F^{(-)}_{\mu \nu}F^{(-)\mu \alpha}F^{(-)\nu}{}_{\alpha}
         +F^{(+)}_{\mu \nu}F^{(+)\mu\alpha}F^{(+)\nu}{}_{\alpha} \Big)
\nonumber
\\
&&+ \frac{1}{540}\mu^4\Big[ 41\Big( \big( D^{(-)}_{\mu \nu \alpha} \big)^2
    +\big( D^{(+)}_{\mu \nu \alpha} \big)^2 \Big)
  - 10 \Big( \big( D^{(-)\mu}{}_{\mu \alpha}\big)^2
            +\big( D^{(+)\mu}{}_{\mu \alpha}\big)^2 \Big) \Big] \Big\}\,.
\label{append3}
\end{eqnarray}

%------------------------------------------------------------------------

\newpage
\begin{center}
              \Large Figure Caption \large
\end{center}
Fig.1. MARK--II \cite{MARK-II} cross section data for
$\gamma \gamma \rightarrow \pi^+ \pi^-$ for CMS production angles
$|cos\theta |\leq 0.6$.
   The experimental points in the region $m_{\pi\pi}<$0.5 GeV were
only included in the analysis.
   The dotted line shows the QED Born contribution; the dashed and
dash-dotted lines show the results of the successive inclusion
of $p^4$--contributions and one--loop corrections.
   Both lines are calculated with
$(\widetilde{L}_9 + \widetilde{L}_{10})=4.2 \cdot 10^{-3}$,
corresponding to the fit of the total cross section data together
with the parameters of Table 1.
   The solid line corresponds to the direct fit of the experimental points for
$m_{\pi\pi}<$0.5 GeV without including the experimental parameters of
Table 1.

\end{document}